\documentclass[%
 reprint,
superscriptaddress,
 amsmath,amssymb,
 aps,
pra,
]{revtex4-1}

\usepackage{graphicx}% Include figure files
\usepackage{dcolumn}% Align table columns on decimal point
\usepackage{bm}% bold math
\usepackage{hyperref}% add hypertext capabilities
\usepackage[version=4]{mhchem}

\begin{document}

\title{Space qualification of ultrafast laser written integrated waveguide optics}

\author{Simone Piacentini}
\thanks{These authors contributed equally to this work.}
\affiliation{Dipartimento di Fisica, Politecnico di Milano, Piazza Leonardo da Vinci 32, 20133 Milano, Italy}
\affiliation{Istituto di Fotonica e Nanotecnologie, Consiglio Nazionale delle Ricerche (IFN-CNR), Piazza Leonardo da Vinci 32, 20133 Milano, Italy}
\author{Tobias Vogl}
\thanks{These authors contributed equally to this work.}
\affiliation{Centre for Quantum Computation and Communication Technology, Department of Quantum Science, Research School of Physics and Engineering, The Australian National University, Acton ACT 2601, Australia}
\affiliation{Institute of Applied Physics, Abbe Center of Photonics, Friedrich-Schiller-Universität Jena, 07745 Jena, Germany}
\affiliation{Cavendish Laboratory, University of Cambridge, JJ Thomson Avenue, Cambridge CB3 0HE, United Kingdom\\
\vspace{0.3cm}
$\textnormal{Corresponding authors: tobias.vogl@uni-jena.de and giacomo.corrielli@polimi.it}$}
\author{Giacomo Corrielli}
\affiliation{Dipartimento di Fisica, Politecnico di Milano, Piazza Leonardo da Vinci 32, 20133 Milano, Italy}
\affiliation{Istituto di Fotonica e Nanotecnologie, Consiglio Nazionale delle Ricerche (IFN-CNR), Piazza Leonardo da Vinci 32, 20133 Milano, Italy}
\author{Ping Koy Lam}
\affiliation{Centre for Quantum Computation and Communication Technology, Department of Quantum Science, Research School of Physics and Engineering, The Australian National University, Acton ACT 2601, Australia}
\author{Roberto Osellame}
\affiliation{Dipartimento di Fisica, Politecnico di Milano, Piazza Leonardo da Vinci 32, 20133 Milano, Italy}
\affiliation{Istituto di Fotonica e Nanotecnologie, Consiglio Nazionale delle Ricerche (IFN-CNR), Piazza Leonardo da Vinci 32, 20133 Milano, Italy}

\date{\today}

\begin{abstract}
Satellite-based quantum technologies represent a possible route for extending the achievable range of quantum communication, allowing the construction of worldwide quantum networks without quantum repeaters. In space missions, however, the volume available for the instrumentation is limited, and footprint is a crucial specification of the devices that can be employed. Integrated optics could be highly beneficial in this sense, as it allows for the miniaturization of different functionalities in small and monolithic photonic circuits. In this work, we report on the qualification of waveguides fabricated in glass by femtosecond laser micromachining for their use in a low Earth orbit space environment. In particular, we exposed different laser written integrated devices, such as straight waveguides, directional couplers, and Mach-Zehnder interferometers, to suitable proton and $\gamma$-ray irradiation. Our experiments show that no significant changes have been induced to their characteristics and performances by the radiation exposure. Our results, combined with the high compatibility of laser-written optical circuits to quantum communication applications, pave the way for the use of laser-written integrated photonic components in future satellite missions.
\end{abstract}

\maketitle

\section{Introduction}
Optical quantum technologies will revolutionize future information processing, communication, and sensing applications \cite{10.1088/1367-2630/aad1ea}. Global quantum networks for example can transform the algorithmic security of the internet into a measurable physical security, without the need for technical, institutional, or political trusted intermediates \cite{Wehnereaam9288}. Such quantum networks rely on the distribution of entangled photon pairs or single-photons over free-space or fiber links. Owing to the strong optical absorption of glass, which scales exponentially with the light propagation length, the communication distance using fibers is typically limited to a few hundreds of kilometers \cite{Inagaki:13,PhysRevLett.121.190502}. Without quantum repeaters, which do not exist yet, fiber-based quantum communication is thus limited to metropolitan networks. To overcome these limitations, the possibility of using relay satellites has been explored in the past years. There have been demonstrations of spontaneous parametric down-conversion in space \cite{PhysRevApplied.5.054022}, satellite-to-ground quantum-limited communication \cite{10.1038/nphoton.2017.107}, quantum-limited measurements of optical signals from a geostationary satellite for coherent communication \cite{Gunthner:17}, and the retro-reflection of single-photon states from satellites to characterize space-to-ground quantum links \cite{Yin:13,PhysRevLett.115.040502,PhysRevLett.116.253601}. The current state-of-the-art is the \textit{Micius} satellite, which performed entanglement distribution \cite{Yin1140}, quantum teleportation \cite{Ren2017}, and a quantum key exchange \cite{Liao2017} over 1200 km, and even on an intercontinental scale between Europe and China \cite{PhysRevLett.120.030501}. Furthermore, with the large variety of proposals and funded missions \cite{10.1016/j.actaastro.2007.12.039,10.1117/12.2041693,10.1117/12.2067548,Oi2017,Kerstel2018,Haber2018QubeA}, there are many more satellites and space experiments to be launched in the next few years.\\
\indent Regarding ground-based photonic quantum technologies, the current research trend is that of integrating all functionalities, i.e. photon creation, manipulation and detection, into miniaturized photonic chips \cite{sansoni2017two,orieux2013direct,qiang2018large,carolan2015universal,najafi2015chip}. In fact, waveguide-based architectures have the twofold advantage over bulk-optics of reducing substantially the footprint of the devices, while guaranteeing a unique degree of interferometric stability of the light paths. Given these advantages, many of the near-future quantum space missions will rely on, or could at least benefit from such integrated waveguide optics, as they allow for much more compact payloads. Integrated optics is of particular importance in space missions where device compactness is a crucial requirement. Among the various integrated platforms tested in the past decade, optical waveguides produced by Ultrafast Laser Micromachining (ULM) in glass proved to be a valuable tool for quantum technology development \cite{marshall2009laser,grafe2014chip,atzeni2018integrated,polino2019experimental,anton2019interfacing}. This fabrication technique consists in inducing a localized refractive index increase in the glass substrate by the direct in-depth focussing of a femtosecond-pulsed laser beam \cite{della2008micromachining}. Optical waveguides fabricated in this way show high guiding performances both in the visible and in the near-infrared range, up to the telecommunication C-band, and exhibit a very good connectivity with both optical fibers and free-space coupling. In addition, laser-written waveguides can be tailored for showing a low degree of birefringence, on the order of $10^{-6}$ \cite{corrielli2018}, and are capable of supporting the polarization encoding of quantum information, widely exploited in free-space quantum links \cite{vest2014design}. All these factors contribute in making laser written optical circuits highly appealing for space-based quantum communication applications, especially as quantum light sources could also be directly integrated and interfaced with the waveguides \cite{0022-3727-50-29-295101}.\\
\indent A necessary requirement for any component to be used in a space scenario is the certification and validation in space environments \cite{GSFC-STD-7000}. This includes thermal and vacuum cycling, shock and vibration tests, as well as radiation hardening. Some key components for quantum optics have been miniaturized and space qualified already, for example single-photon detectors \cite{Tan:13} and single-photon sources \cite{10.1038/s41467-019-09219-5,doi:10.1021/acsphotonics.9b00314}. Of particular interest are radiation effects, since there is usually not much knowledge about the interaction of high energy radiation and the component. Furthermore, testing this aspect requires special sites with access to particle accelerators and $\gamma$-ray sources.\\
\indent Here, we report on the space qualification of ultrafast laser written integrated waveguide optics. We model radiation environments for relevant orbits and mission lengths using the SPace ENVironment Information System (SPENVIS), provided by the European Space Agency \cite{spenvis}. We then expose various waveguide components, such as straight waveguides, directional couplers, and Mach-Zehnder interferometers, to protons and $\gamma$-rays. We investigate combined effects, as well as effects from isolated irradiation types. All components are designed and optimized for telecom wavelengths, either for 850 nm or 1550 nm, but the platform can be generalized to other wavelengths as well. We also carry out thermal and vacuum cycling.

\section{Space environments}
\begin{figure}[hbt]
\includegraphics[width=\linewidth]{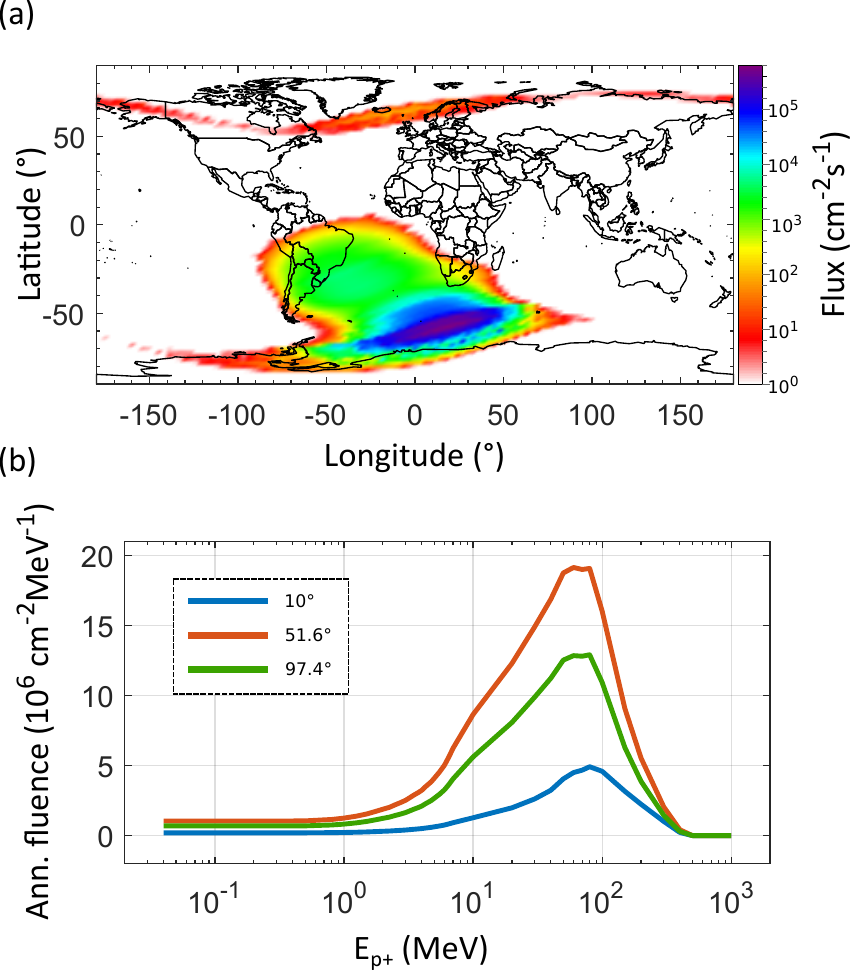}
\caption{Radiation levels in space. (a) Geographical distribution of the proton flux at an altitude of 700 km. The flux was calculated with the AP-8 MAX model in SPENVIS. (b) Integrated annual fluence spectra after 1.85 mm of aluminium for important spacecraft trajectories:  near-equatorial orbit of the Van Allen Probes A and B (10$^\circ$ OI), orbit of the International Space Station (51.6$^\circ$ OI), and near-polar orbit of the \textit{Micius} satellite (97.4$^\circ$ OI). The altitude for all trajectories is 700 km.} 
\label{fig1}
\end{figure}
\begin{figure*}[hbt]
\centerline{\includegraphics[width=\linewidth]{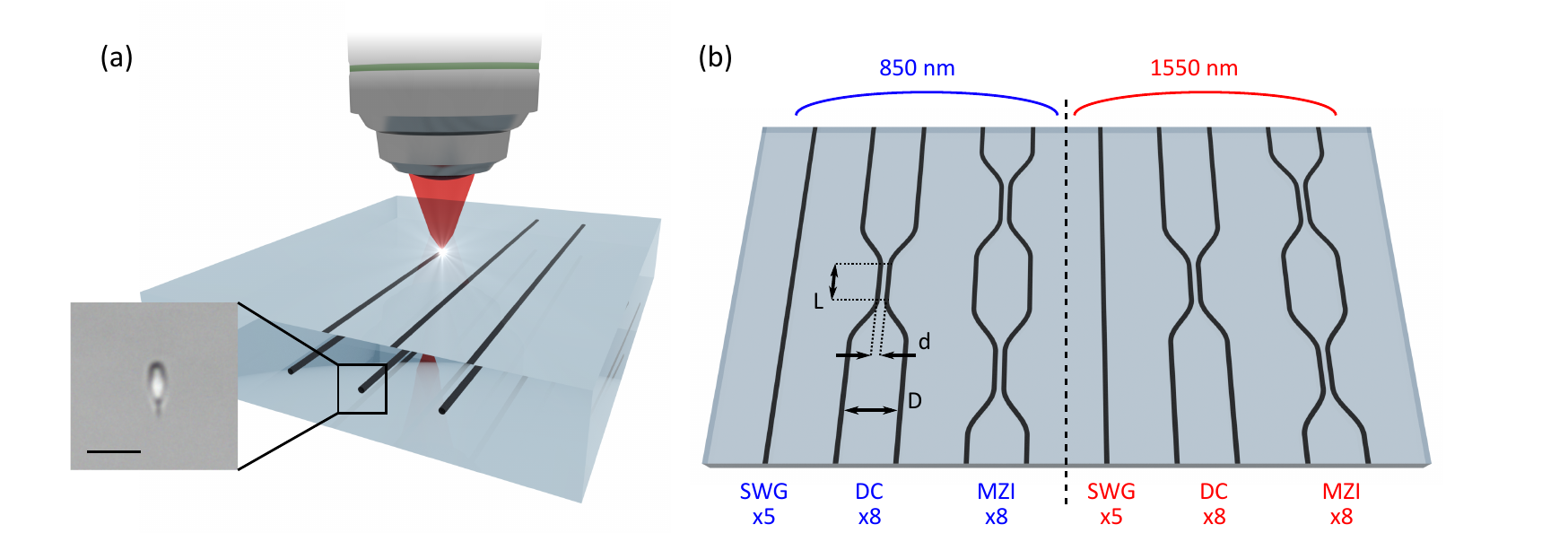}}
\caption{(a) Pictorial representation of the ULM process. A femtosecond laser beam is focused in the bulk of a glass substrate, for producing high quality optical waveguides. In the inset, a microscope picture of the cross section of a waveguide operating at 850 nm is shown. The scale bar indicates 10 $\mu$m. (b) Schematic of the content of the 7 fabricated samples. Each of them contains a set of 5 SWGs, 8 DCs and 8 MZIs, operating both at 850 nm and 1550 nm wavelength. $L$ indicates the coupling length of the DCs, $d$ represents their coupling distance while $D$ indicates the distance between the DCs ports.} 
\label{fig2}
\end{figure*}
The geomagnetic field shields the Earth from cosmic rays and solar winds and thus provides necessary protection for all life on Earth. The magnetic field thereby traps the charged cosmic particles in the Van-Allen belts. As the inner radiation belt can extend down to altitudes below 200 km near the poles, the trapped radiation causes a harmful risk for any satellite orbiting through the stream of particles. As shown in figure \ref{fig1}(a), the total irradiation dose is non-isotropically distributed in the Low Earth Orbit (LEO), making any required shielding dependent on the spacecraft's trajectory or Orbital Inclination (OI). Due to the misalignment of Earth's rotation and magnetic dipole axis, the highest radiation flux appears near the South Atlantic. We choose a representative altitude of 700 km for all following simulations, which is higher than any expected orbit for potential space quantum communication missions. Therefore, our space-certification is valid for all missions with altitudes below 700 km. The annual fluence spectrum for protons after a typical aluminium shield with a thickness of 1.85 mm is shown in figure \ref{fig1}(b), where the proton energies range from about 100 keV up to 400 MeV. While particles with low kinetic energy are absorbed by the Al shield, particles with higher energy loose at least some of their energy while passing through matter. This leads to a finite flux of low-energy particles in the energy spectrum. It is worth noting that charge carriers with a lower kinetic energy are expected to be more harmful, as they can deposit more energy into the waveguide due to their higher interaction cross section (see Supplementary Material). Depending on orbital inclination, the integrated annual proton fluence amounts up to $2.9\times 10^{9}$ cm$^{-2}$, so for a typical mission length of three years the fluence would be less than $10^{10}$ cm$^{-2}$. The total absorbed ionizing dose is on the order of a few 10 Gy per year.

\section{Device fabrication and irradiation}
The waveguide inscription was performed by focussing a femtosecond laser beam, produced by an homemade Yb:KYW cavity-dumped mode-locked oscillator (pulse duration of 300 fs, repetition rate of 1 MHz, wavelength of 1030 nm), inside the bulk of an Eagle XG borosilicate glass substrate (from Corning). The focussing optics used was a microscope objective with 50$\times$ magnification and 0.6 numerical aperture. The waveguides were written by scanning the sample at a speed of 40 mm/s and by performing multiple overlapped scans, employing an air bearing motorized three-axis translation stage (Aerotech FiberGlide 3D). For the fabrication of the waveguides operating at 850 nm we used a pulse energy of 370 nJ and 5 scans, while for the waveguides at 1550 nm wavelength we used a pulse energy of 480 nJ and 6 scans. A pictorial schematic of the ULM process is represented in figure \ref{fig2}(a). After the laser irradiation step, a sample thermal annealing was performed, in order to obtain single mode guidance and to reduce waveguide birefringence \cite{arriola2013,corrielli2018}. This consisted in a heating ramp up to a temperature of 600 $^\circ$C at the rate of 100 $^\circ$C/h, followed by another ramp up to 750 $^\circ$C with a rate of 75 $^\circ$C/h, and finally a cooling to room temperature with a rate of -12 $^\circ$C/h. A microscope picture of the cross section of an 850 nm waveguide after the thermal annealing step is presented in the inset of figure \ref{fig2}(a).\\
\indent In order to explore different conditions of protons and $\gamma$-rays exposure, we fabricated 7 identical samples, each of them containing 2 equal subsets of integrated components each operating at one of the two wavelengths of choice. Each subset is composed by 5 Straight WaveGuides (SWGs), which allow to analyze waveguide losses and birefringence, 8 Directional Couplers (DCs), for observing possible effects of radiation exposure on the evanescent coupling, and 8 Mach-Zehnder Interferometers (MZIs), for studying the stability of the waveguide optical paths. A schematic of the sample content is sketched in figure \ref{fig2}(b). The DCs operating at 850 nm were fabricated with a fixed interaction distance $d$ = 8 $\mu$m and an interaction length $L$ ranging from 1 mm to 4.5 mm with a step of 0.5 mm, while the DCs for 1550 nm were fabricated with $d$ = 9 $\mu$m and $L$ from 0 mm to 3.5 mm with the same step. The separation of the DCs arms outside the coupling region was set to $D$ = 60 $\mu$m. The same geometrical configurations were used for the couplers composing the MZIs. In addition, an optical phase difference of $\pi$/2 was implemented in all MZIs by a geometrical deformation of one of the two arms, for enhancing their response to possible optical phase shifts induced by the protons and $\gamma$-ray exposure. A constant radius of curvature $R$ = 50 mm was used for fabricating the S-bends of all devices. The inscription depth was varied in the following way: 5 samples were written 170 $\mu$m beneath the substrate top surface, while the other 2 were written at the depth of 200 $\mu$m. Subsequently, after the annealing step, the samples were immersed in a 10\,\% aqueous solution of hydrofluoric acid at 35 $^\circ$C for 100 minutes, for reducing the depth of the devices to the desired values of 40 $\mu$m and 10 $\mu$m. These values were chosen to match the projected range of protons with 770 keV kinetic energy in case of the 10 $\mu$m depth (see Supplementary Material). Protons with that energy are thus injected directly into the waveguiding structure. Higher energy protons with 3 MeV pass through the 10 $\mu$m and 40 $\mu$m deep waveguide. It is worth highlighting that the direct inscription of the waveguides very close to surface (e.g. 10 $\mu$m) poses additional experimental challenges, mainly arising from the interaction with the fs laser beam and the substrate surface, which would have limited the achievable devices performances. Lastly, the waveguides were polished to optical quality at the lateral facets, for allowing efficient light coupling in the waveguides by butt coupling with optical fibers.\\
\indent After a preliminary optical characterization, that will be discussed in the next paragraph, we exposed the devices to protons and $\gamma$-rays. In particular, among the 7 fabricated samples, six have been exposed to different irradiation conditions, and one was kept as control sample, without any exposure. For the proton irradiation we have used a 1.7 MV tandem accelerator, thus allowing us to use kinetic ion energies up to 3.4 MeV per charge number. The initial target for the ion sputter source was \ce{TiH}. The \ce{Ti} ions, however, have been filtered out by a 90$^\circ$ magnet. The kinetic energy of the ions is well defined, typically within $\pm 5$ keV, while the target fluence can be reached with an accuracy better than $\pm 10$\,\%. To avoid ion channeling along the principal crystallographic directions, the angle of incidence of the particles was 7$^\circ$. Otherwise the protons are steered by electrostatic interaction with ion columns in the target material with reduced energy loss. As a consequence, the projected range of the protons in matter can be significantly larger than predicted by random collision models (see Supplementary Material). Tilting the target by 7$^\circ$ minimizes this effect and is thus standard in particle physics. It should be noted that this is likely not necessary in an amorphous material like glass, but the beam path of our particle accelerator was aligned to the standard 7$^\circ$ angle of incidence. The irradiation was carried out under high vacuum with a pressure of $10^{-7}$ Torr and at room temperature. Coincidentally, this pressure is in the range of the atmospheric pressure in low-Earth orbit. While we did not test the samples \textit{in-situ} under vacuum, we can nevertheless conclude that high vacuum does not alter any of the characteristics, as will be shown in the following section. This is of course expected, as borosilicate glass is compatible for ultra-high vacuum and is often used as view ports for vacuum chambers.\\
\indent To study theoretically the interaction of the protons with the glass, we used the SRIM (Stopping and Range of Ions in Matter) code \cite{10.1016/j.nimb.2010.02.091}. With these simulations (see Supplementary Material), we predict that protons with a kinetic energy of 770 keV are stopped at the actual waveguide depth of 10 $\mu$m, given the composition of Eagle XG glass. Furthermore, we predict that protons with a kinetic energy of 3 MeV will pass through all waveguides and deeply penetrate into the glass. Hence, we study the effects of protons being injected directly into the waveguiding regions of the glass as well as effects determined by collisions of high-energy protons passing through the glass. We tested a proton fluence of $10^{10}$ cm$^{-2}$, as required for the three year long mission (see SPENVIS simulation), as well as with a 100-fold increased fluence of $10^{12}$ cm$^{-2}$.\\
\indent For the $\gamma$-irradiation, we used the radioactive isotope Co-60, which emits two $\gamma$-rays with 1.17 and 1.33 MeV energy per decay with a branching ratio of nearly 100\,\% \cite{toi}. During the irradiation, the samples were placed between solid water to obtain the same conditions as during the dose rate calibration with an ionisation chamber. The dose rate at the sample was 4.329(36) Gy/min. The details of the irradiation parameters used for every sample, as well as the corresponding inscription depth, are summarized in table \ref{tab:samples}. It is worth noting that we have irradiated one sample (\# 5) with both protons and $\gamma$-rays to study combined radiation effects.

\begin{table}[htbp]
\centering
\caption{Summary of the protons (p$^+$) and $\gamma$-rays exposure conditions for the different samples. Sample 1 was not irradiated and kept as a reference.}
\label{tab:samples}
\begin{ruledtabular}
\begin{tabular}{ccccc}
%\hline
Sample &  Depth & p$^+$ energy & p$^+$ fluence & $\gamma$-rays dose \\
$\#$ & ($\mu$m) & (MeV) & (cm$^{-2}$) & (Gy) \\
\colrule
1 & 10 & - & - & - \\
2 & 10 & - & - & 10.0(11) \\
3 & 10 & - & - & 100.0(24) \\
4 & 10 & 0.77 & 10$^{12}$ & - \\
5 & 10 & 3 & 10$^{12}$ & 50.0(15) \\
6 & 40 & 3 & 10$^{12}$ & - \\
7 & 40 & 3 & 10$^{10}$ & - \\
%\hline
\end{tabular}
\end{ruledtabular}
\end{table}

\section{Comparison of the results}
\begin{figure*}[hbt]
\centerline{\includegraphics[width=\linewidth]{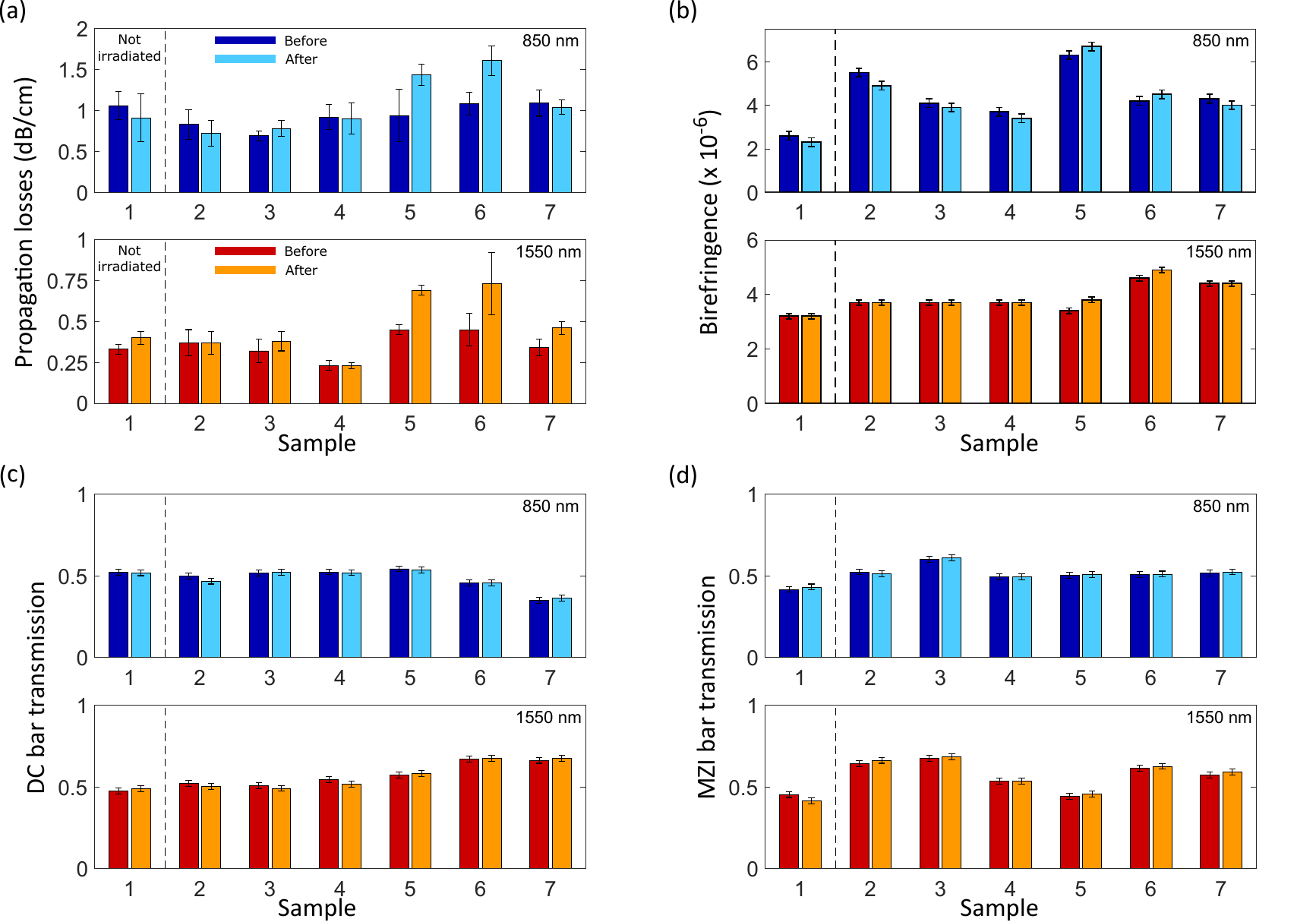}}
\caption{Comparison of the optical properties of the fabricated devices before and after the radiation exposure. In all panels, sample 1 indicates the reference sample, which was not irradiated. Darker (lighter) columns indicate the data taken before (after) irradiation.  The top and the bottom subpanels refer to the devices operating at 850 nm and 1550 nm respectively. (a) Measurement of the waveguides propagation losses. (b) Measurement of the waveguides birefringence. (c) Values of bar transmission measured on the directional couplers showing the splitting ratio close to 50:50 for every sample, using vertically polarized light. (d) Same measurement as panel (c), but performed on the Mach-Zehnder interferometers.} 
\label{fig3}
\end{figure*}

A detailed inspection of the optical properties of all fabricated devices has been performed before and after their exposure to protons and $\gamma$-rays. These include waveguide propagation losses, waveguide birefringence, and the splitting performances of the DCs and the MZIs. A temporal interval of approximately 3 months passed between the two characterizations. Unless otherwise specified, cleaved optical fibers were used for the light coupling in the waveguides (model SM800 for 850 nm light and SMF28 for 1550 nm light), while a microscope objective (0.5 NA) was used for the light collection at the waveguide output. We report in the following a comparison of the results of the two sets of measurements performed on all fabricated devices. 

\subsection{Propagation losses}
The values of Propagation Losses ($PL$) were obtained by first measuring the total insertion losses $IL_{dB}=-10\,\mathrm{Log}(P_{IN}/P_{OUT})$ of the SWGs, where $P_{IN}$ and $P_{OUT}$ represent the optical power measured at the waveguide input and output respectively. From these values we then subtracted the reflection losses of 0.17 dB arising from the glass/air interface at the waveguide output, and the Coupling Losses ($CL$) contribution caused by the mode mismatch between the fibers and the waveguides. We experimentally measured the values of $CL$ by acquiring with a CCD camera the near field intensity profiles $I_{F}(x,y)$ and $I_{WG}(x,y)$ of the fibers and the waveguide modes respectively, and by numerically computing their superposition integral $\eta$, as defined in \cite{osellame2004optical}. The value of $CL$ is then calculated as $-10\, \mathrm{Log}(\eta)$. Finally, the value of $PL$ is obtained by dividing the result by the total length of the SWGs, which is 2.2 cm. This measurement was performed on all fabricated SWGs.\\
\indent In Figure \ref{fig3}(a) we report the $PL$ values calculated by averaging the results obtained from the whole set of SWGs at a specific wavelength, for each sample. The error bars indicate their standard deviation. Darker colors represent results measured before irradiation while lighter colors refer to the results obtained after irradiation. From the values obtained on the reference sample it can be seen that this measurement is affected by a certain degree of variability, which we mainly attribute to the difficulty in achieving the same experimental $CL$ in different measurements performed on the same SWG. In fact, the values of $CL$ estimated numerically are the lowest possible, ideally due only to mode mismatch, while, experimentally, we can have an additional misalignment between the fiber and the waveguide that produce higher experimental $CL$. In light of this it can be seen that, within this experimental uncertainty, the values of $PL$ measured on samples 2, 3, 4 and 7 did not change significantly after the radiation exposure. On the contrary, samples 5 and 6 showed a statistically relevant loss increase up to 0.5 dB/cm. The degradation of these samples is further confirmed by an optical microscope inspection of their cross sections, which showed a reduced brightness of their core in transmission imaging (see Supplementary Material). This analysis shows that the irradiation with $\gamma$-rays alone (samples 2 and 3) did not affect the waveguide losses. The same holds for the irradiation with either a high-fluence (10$^{-12}$ cm$^{-2}$) of low-energy protons (770 keV, sample 4) or a low-fluence (10$^{-10}$ cm$^{-2}$) of more energetic protons (3 MeV, sample 7). Waveguide alterations, instead, are caused by the combined effects of high-energy and high-fluence protons, which possibly create a significant number of defects and vacancies in the glass lattice during their passage, leading to additional light scattering. A deeper understanding of the waveguide alteration mechanisms would require further experimental investigation, and it goes beyond the scope of this work. One should keep in mind that the proton irradiation is 100 times above what is required for our space qualification. If the waveguides are intended to be used in such heavy irradiation environments, a different shielding material could be adopted. For example, a tantalum aluminium alloy can achieve higher ion absorption with the same weight compared to pure aluminium.

\subsection{Birefringence}
We performed the birefringence measurement by injecting light in a SWG with a known polarization state and then measuring the Stokes vector of the light at the waveguide output. The preparation of the polarization state was performed by a fixed linear polarizer and a pair of rotating half- and quarter-waveplates, and, in this case, light was coupled to the waveguide by means of free-space focussing on its input facet. The measurement of the Stokes parameter, instead, was performed with a rotating quarter-waveplate and a rotating linear polarizer. All rotating elements were mounted on motorized rotators (Thorlabs, PRM1Z8 model). This measurement was repeated several times for each SWG, by changing the polarization state at the input. This allowed to experimentally measure the M\"uller transformation associated to the waveguide. The birefringence value was then obtained by fitting the measured matrix with the model of a generic linear polarization retarder. The uncertainty associated to this measurement was estimated to be $2\cdot10^{-7}$ for the data at 850 nm and $1\cdot10^{-7}$ for the data taken at 1550 nm. This estimation has been performed by measuring a repeated number of times the residual polarization retardation of the empty setup, which arises from the non-ideal behavior of the polarization components used, and the finite reproducibility of their reciprocal alignment.\\
\indent The birefringence measurement was performed on two SWGs (one for each wavelength subset) per sample. The comparison between the data obtained before and after the irradiation are shown in Fig. \ref{fig3}(b), where it is possible to appreciate that no significant changes in waveguides birefringence occurred within the experimental precision of our measurement.

\subsection{DCs and MZIs analysis}
For analyzing the splitting ratio of the DCs and the MZIs, we coupled light in one of the two input ports, and then we collected the optical powers $P_{\mathrm{bar}}$ and $P_{\mathrm{cross}}$ transmitted at the bar and cross output ports respectively. We then computed the value of the normalized bar transmission of each device, defined as $T=P_{\mathrm{bar}}/(P_{\mathrm{bar}}+P_{\mathrm{cross}})$. This measurement has been performed on every DC and MZI of all samples, using both horizontally and vertically polarized light at the input. The experimental uncertainty associated to this measurement is estimated to be $\approx$ 1 \%, and we mainly attribute this to the effect of background light not coupled into the devices.\\ 
\indent In all samples we did not observe any significant variation in the transmission of the DCs and the MZIs before and after the exposure to the protons and the $\gamma$-rays. As a representative evidence of this fact, we plot in figures \ref{fig3}(c) and \ref{fig3}(d) the $T$ values measured for the DCs and the MZIs that, in each sample, showed the splitting ratio closer to 50 \%. These are, in fact, devices whose performances are most sensitive to possible variations in the evanescent coupling coefficient and, in the case of the MZIs, in a change of the internal optical path difference. The average variation of splitting ratio measured for these devices is 1.1 \%, which is comparable to our experimental uncertainty. For the sake of clarity, we have reported in figures \ref{fig3}(c) and \ref{fig3}(d) only the values obtained for vertically polarized light. However, very similar results have been obtained also for the horizontal polarization case. The complete data set, comprising all values measured on all DCs and MZIs with both polarizations, is available in the Supplementary Material.

\section{Temperature insensitivity}
\begin{figure}[hbt]
\includegraphics[width=\linewidth]{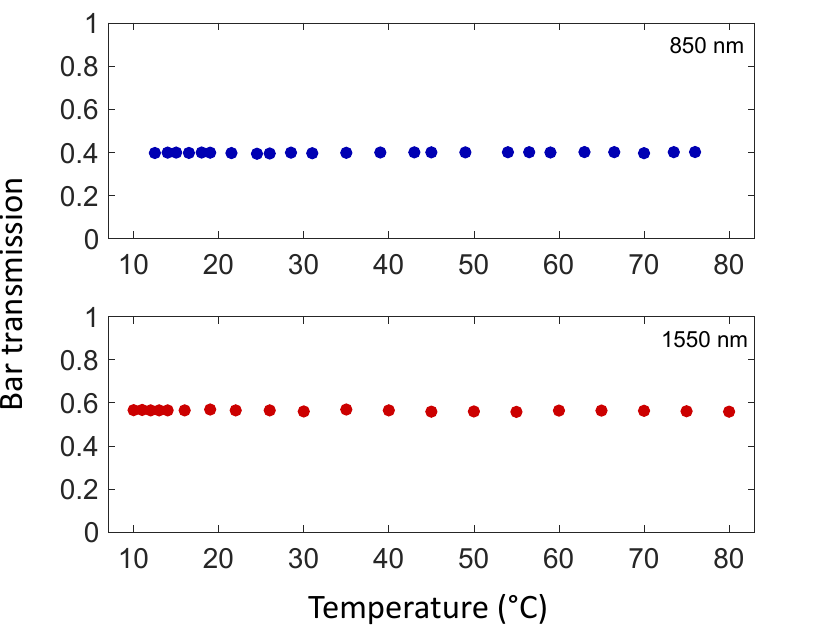}
\caption{Qualification of the laser written MZIs functioning with respect to temperature variations. (a) Measurement performed on a MZI from sample 1, operating at 850 nm. (b) Measurement performed on a MZI from sample 4, operating at 1550 nm. Experimental error bars are smaller than data markers.} 
\label{fig4}
\end{figure}

\indent As a final analysis, we studied the performances of the laser written devices with respect to variations of the sample temperature. In fact, previous space missions reported moderate temperature fluctuations ($\Delta T=20$ $^\circ$ C) on a satellite with a sun-synchronous orbit at an altitude of 630 km and inclination of 98$^\circ$ \cite{doi:10.1029/2009JA014699}. The higher temperature limit is reached when the satellite is in sunlight, while the lower temperature limit is reached when the satellite is on the dark side of the Earth. Due to this reason, we validate the stability of our devices across a temperature range of $\Delta T=70$ $^\circ$C. The measurement was performed by controlling the sample temperature with a Peltier plate, driven by a voltage generator, in order to induce either a negative or a positive temperature difference with respect to room conditions ($\approx$ 25 $^\circ$C). The sample was mounted in a copper enclosure (in direct contact to the Peltier plate) for guaranteeing a uniform temperature distribution in the whole glass. The analyzed temperature range spans between 10 $^\circ$C and 80 $^\circ$C.\\
\indent We performed the measurement on two MZIs, as these devices allow to test simultaneously the stability of the DC splitting ratio and the stability of the optical path differences. In particular, we have decided to test one MZI from sample 1 (thus not exposed to any radiation) operating at 850 nm, and one MZI from sample 4 operating at 1550 nm. In both cases we have chosen the one with the bar transmission value closer to 50 \%. The results are shown in Fig. \ref{fig4}(a) and Fig. \ref{fig4}(b) respectively. In this case, error bars are smaller than data markers. From the two graphs it is possible to conclude that in both cases the bar transmission values remain stable across the whole temperature range.

\section{Discussion and conclusion}
\indent In this work, we have performed the space qualification of laser-written waveguides, operating at two relevant telecommunication wavelengths (850 and 1550 nm) and fabricated in borosilicate glass, for their use in a LEO space environment. In particular, we fabricated 294 integrated devices, exposed them to protons and $\gamma$-rays irradiation, and experimentally verified that their properties were largely unaffected by these processes. In doing so, we monitored different performances such as propagation losses, waveguide birefringence, evanescent coupling, and stability of interferometric waveguide paths. The exposure parameters and the device inscription depths were chosen to simulate the typical space environment at the altitude of 700 km from the Earth surface. Among all monitored performances, we have observed a waveguide degradation only in terms of propagation losses increase, solely for the waveguides irradiated with the highest proton fluence, 100 times stronger than what expected in the typical space environment considered for this space qualification. In addition to the resistance against radiations, we also verified that the functioning of laser written devices in borosilicate glass is largely independent of the environment temperature, within a tested range of 70 $^\circ$C. Moreover, during our extensive tests the samples have been exposed to high vacuum without any observable degradation.\\
\indent Our results pave the way to the employment of laser-written photonic circuits in satellite-based experiments. This will be highly beneficial to photonic quantum communication, for which satellites represent a promising route for overcoming the distance limitations arising from losses in optical fibers. Beyond quantum technologies, we expect that our results will have an important impact also in perspective of space-based astronomical observations, for which complex on-chip interferometry is an important resource \cite{labadie2016astronomical}, and where laser-written photonic circuits represent a crucial enabling technology \cite{diener2017towards,norris2020first}.

\section*{Funding Information}
This work was funded by the Australian Research Council (CE170100012, FL150100019) and by the European Research Council (ERC) under the European Union’s Horizon 2020 research and innovation programme (grant agreement 742745 - www.capable-erc.eu).

\section*{Acknowledgments}
We acknowledge access to the Heavy Ion Accelerator Capability, part of the NCRIS facilities at the Australian National University. We also acknowledge the support of the ANSTO, in providing the gamma irradiation facility used in this work.

\section*{Disclosures}
The authors declare no conflicts of interest.

\section*{Supplementary Materials} See \href{link}{Supplementary Materials} for supporting content.


\begin{thebibliography}{47}%
\makeatletter
\providecommand \@ifxundefined [1]{%
 \@ifx{#1\undefined}
}%
\providecommand \@ifnum [1]{%
 \ifnum #1\expandafter \@firstoftwo
 \else \expandafter \@secondoftwo
 \fi
}%
\providecommand \@ifx [1]{%
 \ifx #1\expandafter \@firstoftwo
 \else \expandafter \@secondoftwo
 \fi
}%
\providecommand \natexlab [1]{#1}%
\providecommand \enquote  [1]{``#1''}%
\providecommand \bibnamefont  [1]{#1}%
\providecommand \bibfnamefont [1]{#1}%
\providecommand \citenamefont [1]{#1}%
\providecommand \href@noop [0]{\@secondoftwo}%
\providecommand \href [0]{\begingroup \@sanitize@url \@href}%
\providecommand \@href[1]{\@@startlink{#1}\@@href}%
\providecommand \@@href[1]{\endgroup#1\@@endlink}%
\providecommand \@sanitize@url [0]{\catcode `\\12\catcode `\$12\catcode
  `\&12\catcode `\#12\catcode `\^12\catcode `\_12\catcode `\%12\relax}%
\providecommand \@@startlink[1]{}%
\providecommand \@@endlink[0]{}%
\providecommand \url  [0]{\begingroup\@sanitize@url \@url }%
\providecommand \@url [1]{\endgroup\@href {#1}{\urlprefix }}%
\providecommand \urlprefix  [0]{URL }%
\providecommand \Eprint [0]{\href }%
\providecommand \doibase [0]{https://doi.org/}%
\providecommand \selectlanguage [0]{\@gobble}%
\providecommand \bibinfo  [0]{\@secondoftwo}%
\providecommand \bibfield  [0]{\@secondoftwo}%
\providecommand \translation [1]{[#1]}%
\providecommand \BibitemOpen [0]{}%
\providecommand \bibitemStop [0]{}%
\providecommand \bibitemNoStop [0]{.\EOS\space}%
\providecommand \EOS [0]{\spacefactor3000\relax}%
\providecommand \BibitemShut  [1]{\csname bibitem#1\endcsname}%
\let\auto@bib@innerbib\@empty
%</preamble>
\bibitem [{\citenamefont {Ac{\'{\i}}n}\ \emph {et~al.}(2018)\citenamefont
  {Ac{\'{\i}}n}, \citenamefont {Bloch}, \citenamefont {Buhrman}, \citenamefont
  {Calarco}, \citenamefont {Eichler}, \citenamefont {Eisert}, \citenamefont
  {Esteve}, \citenamefont {Gisin}, \citenamefont {Glaser}, \citenamefont
  {Jelezko}, \citenamefont {Kuhr}, \citenamefont {Lewenstein}, \citenamefont
  {Riedel}, \citenamefont {Schmidt}, \citenamefont {Thew}, \citenamefont
  {Wallraff}, \citenamefont {Walmsley},\ and\ \citenamefont
  {Wilhelm}}]{10.1088/1367-2630/aad1ea}%
  \BibitemOpen
  \bibfield  {author} {\bibinfo {author} {\bibfnamefont {A.}~\bibnamefont
  {Ac{\'{\i}}n}}, \bibinfo {author} {\bibfnamefont {I.}~\bibnamefont {Bloch}},
  \bibinfo {author} {\bibfnamefont {H.}~\bibnamefont {Buhrman}}, \bibinfo
  {author} {\bibfnamefont {T.}~\bibnamefont {Calarco}}, \bibinfo {author}
  {\bibfnamefont {C.}~\bibnamefont {Eichler}}, \bibinfo {author} {\bibfnamefont
  {J.}~\bibnamefont {Eisert}}, \bibinfo {author} {\bibfnamefont
  {D.}~\bibnamefont {Esteve}}, \bibinfo {author} {\bibfnamefont
  {N.}~\bibnamefont {Gisin}}, \bibinfo {author} {\bibfnamefont {S.~J.}\
  \bibnamefont {Glaser}}, \bibinfo {author} {\bibfnamefont {F.}~\bibnamefont
  {Jelezko}}, \bibinfo {author} {\bibfnamefont {S.}~\bibnamefont {Kuhr}},
  \bibinfo {author} {\bibfnamefont {M.}~\bibnamefont {Lewenstein}}, \bibinfo
  {author} {\bibfnamefont {M.~F.}\ \bibnamefont {Riedel}}, \bibinfo {author}
  {\bibfnamefont {P.~O.}\ \bibnamefont {Schmidt}}, \bibinfo {author}
  {\bibfnamefont {R.}~\bibnamefont {Thew}}, \bibinfo {author} {\bibfnamefont
  {A.}~\bibnamefont {Wallraff}}, \bibinfo {author} {\bibfnamefont
  {I.}~\bibnamefont {Walmsley}},\ and\ \bibinfo {author} {\bibfnamefont
  {F.~K.}\ \bibnamefont {Wilhelm}},\ }\bibfield  {title} {\bibinfo {title} {The
  quantum technologies roadmap: a european community view},\ }\href
  {https://doi.org/10.1088/1367-2630/aad1ea} {\bibfield  {journal} {\bibinfo
  {journal} {New J. Phys.}\ }\textbf {\bibinfo {volume} {20}},\ \bibinfo
  {pages} {080201} (\bibinfo {year} {2018})}\BibitemShut {NoStop}%
\bibitem [{\citenamefont {Wehner}\ \emph {et~al.}(2018)\citenamefont {Wehner},
  \citenamefont {Elkouss},\ and\ \citenamefont {Hanson}}]{Wehnereaam9288}%
  \BibitemOpen
  \bibfield  {author} {\bibinfo {author} {\bibfnamefont {S.}~\bibnamefont
  {Wehner}}, \bibinfo {author} {\bibfnamefont {D.}~\bibnamefont {Elkouss}},\
  and\ \bibinfo {author} {\bibfnamefont {R.}~\bibnamefont {Hanson}},\
  }\bibfield  {title} {\bibinfo {title} {Quantum internet: A vision for the
  road ahead},\ }\bibfield  {journal} {\bibinfo  {journal} {Science}\ }\textbf
  {\bibinfo {volume} {362}},\ \href {https://doi.org/10.1126/science.aam9288}
  {10.1126/science.aam9288} (\bibinfo {year} {2018})\BibitemShut {NoStop}%
\bibitem [{\citenamefont {Inagaki}\ \emph {et~al.}(2013)\citenamefont
  {Inagaki}, \citenamefont {Matsuda}, \citenamefont {Tadanaga}, \citenamefont
  {Asobe},\ and\ \citenamefont {Takesue}}]{Inagaki:13}%
  \BibitemOpen
  \bibfield  {author} {\bibinfo {author} {\bibfnamefont {T.}~\bibnamefont
  {Inagaki}}, \bibinfo {author} {\bibfnamefont {N.}~\bibnamefont {Matsuda}},
  \bibinfo {author} {\bibfnamefont {O.}~\bibnamefont {Tadanaga}}, \bibinfo
  {author} {\bibfnamefont {M.}~\bibnamefont {Asobe}},\ and\ \bibinfo {author}
  {\bibfnamefont {H.}~\bibnamefont {Takesue}},\ }\bibfield  {title} {\bibinfo
  {title} {Entanglement distribution over 300 km of fiber},\ }\href
  {https://doi.org/10.1364/OE.21.023241} {\bibfield  {journal} {\bibinfo
  {journal} {Opt. Express}\ }\textbf {\bibinfo {volume} {21}},\ \bibinfo
  {pages} {23241} (\bibinfo {year} {2013})}\BibitemShut {NoStop}%
\bibitem [{\citenamefont {Boaron}\ \emph {et~al.}(2018)\citenamefont {Boaron},
  \citenamefont {Boso}, \citenamefont {Rusca}, \citenamefont {Vulliez},
  \citenamefont {Autebert}, \citenamefont {Caloz}, \citenamefont {Perrenoud},
  \citenamefont {Gras}, \citenamefont {Bussi\`eres}, \citenamefont {Li},
  \citenamefont {Nolan}, \citenamefont {Martin},\ and\ \citenamefont
  {Zbinden}}]{PhysRevLett.121.190502}%
  \BibitemOpen
  \bibfield  {author} {\bibinfo {author} {\bibfnamefont {A.}~\bibnamefont
  {Boaron}}, \bibinfo {author} {\bibfnamefont {G.}~\bibnamefont {Boso}},
  \bibinfo {author} {\bibfnamefont {D.}~\bibnamefont {Rusca}}, \bibinfo
  {author} {\bibfnamefont {C.}~\bibnamefont {Vulliez}}, \bibinfo {author}
  {\bibfnamefont {C.}~\bibnamefont {Autebert}}, \bibinfo {author}
  {\bibfnamefont {M.}~\bibnamefont {Caloz}}, \bibinfo {author} {\bibfnamefont
  {M.}~\bibnamefont {Perrenoud}}, \bibinfo {author} {\bibfnamefont
  {G.}~\bibnamefont {Gras}}, \bibinfo {author} {\bibfnamefont {F.}~\bibnamefont
  {Bussi\`eres}}, \bibinfo {author} {\bibfnamefont {M.-J.}\ \bibnamefont {Li}},
  \bibinfo {author} {\bibfnamefont {D.}~\bibnamefont {Nolan}}, \bibinfo
  {author} {\bibfnamefont {A.}~\bibnamefont {Martin}},\ and\ \bibinfo {author}
  {\bibfnamefont {H.}~\bibnamefont {Zbinden}},\ }\bibfield  {title} {\bibinfo
  {title} {Secure quantum key distribution over 421 km of optical fiber},\
  }\href {https://doi.org/10.1103/PhysRevLett.121.190502} {\bibfield  {journal}
  {\bibinfo  {journal} {Phys. Rev. Lett.}\ }\textbf {\bibinfo {volume} {121}},\
  \bibinfo {pages} {190502} (\bibinfo {year} {2018})}\BibitemShut {NoStop}%
\bibitem [{\citenamefont {Tang}\ \emph {et~al.}(2016)\citenamefont {Tang},
  \citenamefont {Chandrasekara}, \citenamefont {Tan}, \citenamefont {Cheng},
  \citenamefont {Sha}, \citenamefont {Hiang}, \citenamefont {Oi},\ and\
  \citenamefont {Ling}}]{PhysRevApplied.5.054022}%
  \BibitemOpen
  \bibfield  {author} {\bibinfo {author} {\bibfnamefont {Z.}~\bibnamefont
  {Tang}}, \bibinfo {author} {\bibfnamefont {R.}~\bibnamefont {Chandrasekara}},
  \bibinfo {author} {\bibfnamefont {Y.~C.}\ \bibnamefont {Tan}}, \bibinfo
  {author} {\bibfnamefont {C.}~\bibnamefont {Cheng}}, \bibinfo {author}
  {\bibfnamefont {L.}~\bibnamefont {Sha}}, \bibinfo {author} {\bibfnamefont
  {G.~C.}\ \bibnamefont {Hiang}}, \bibinfo {author} {\bibfnamefont {D.~K.~L.}\
  \bibnamefont {Oi}},\ and\ \bibinfo {author} {\bibfnamefont {A.}~\bibnamefont
  {Ling}},\ }\bibfield  {title} {\bibinfo {title} {Generation and analysis of
  correlated pairs of photons aboard a nanosatellite},\ }\href
  {https://doi.org/10.1103/PhysRevApplied.5.054022} {\bibfield  {journal}
  {\bibinfo  {journal} {Phys. Rev. Applied}\ }\textbf {\bibinfo {volume} {5}},\
  \bibinfo {pages} {054022} (\bibinfo {year} {2016})}\BibitemShut {NoStop}%
\bibitem [{\citenamefont {Takenaka}\ \emph {et~al.}(2017)\citenamefont
  {Takenaka}, \citenamefont {Carrasco-Casado}, \citenamefont {Fujiwara},
  \citenamefont {Kitamura}, \citenamefont {Sasaki},\ and\ \citenamefont
  {Toyoshima}}]{10.1038/nphoton.2017.107}%
  \BibitemOpen
  \bibfield  {author} {\bibinfo {author} {\bibfnamefont {H.}~\bibnamefont
  {Takenaka}}, \bibinfo {author} {\bibfnamefont {A.}~\bibnamefont
  {Carrasco-Casado}}, \bibinfo {author} {\bibfnamefont {M.}~\bibnamefont
  {Fujiwara}}, \bibinfo {author} {\bibfnamefont {M.}~\bibnamefont {Kitamura}},
  \bibinfo {author} {\bibfnamefont {M.}~\bibnamefont {Sasaki}},\ and\ \bibinfo
  {author} {\bibfnamefont {M.}~\bibnamefont {Toyoshima}},\ }\bibfield  {title}
  {\bibinfo {title} {Satellite-to-ground quantum-limited communication using a
  50-kg-class microsatellite},\ }\href
  {https://doi.org/10.1038/nphoton.2017.107} {\bibfield  {journal} {\bibinfo
  {journal} {Nature Photon.}\ }\textbf {\bibinfo {volume} {11}},\ \bibinfo
  {pages} {502–508} (\bibinfo {year} {2017})}\BibitemShut {NoStop}%
\bibitem [{\citenamefont {G\"{u}nthner}\ \emph {et~al.}(2017)\citenamefont
  {G\"{u}nthner}, \citenamefont {Khan}, \citenamefont {Elser}, \citenamefont
  {Stiller}, \citenamefont {\"{O}mer Bayraktar}, \citenamefont {M\"{u}ller},
  \citenamefont {Saucke}, \citenamefont {Tr\"{o}ndle}, \citenamefont {Heine},
  \citenamefont {Seel}, \citenamefont {Greulich}, \citenamefont {Zech},
  \citenamefont {G\"{u}tlich}, \citenamefont {Philipp-May}, \citenamefont
  {Marquardt},\ and\ \citenamefont {Leuchs}}]{Gunthner:17}%
  \BibitemOpen
  \bibfield  {author} {\bibinfo {author} {\bibfnamefont {K.}~\bibnamefont
  {G\"{u}nthner}}, \bibinfo {author} {\bibfnamefont {I.}~\bibnamefont {Khan}},
  \bibinfo {author} {\bibfnamefont {D.}~\bibnamefont {Elser}}, \bibinfo
  {author} {\bibfnamefont {B.}~\bibnamefont {Stiller}}, \bibinfo {author}
  {\bibnamefont {\"{O}mer Bayraktar}}, \bibinfo {author} {\bibfnamefont
  {C.~R.}\ \bibnamefont {M\"{u}ller}}, \bibinfo {author} {\bibfnamefont
  {K.}~\bibnamefont {Saucke}}, \bibinfo {author} {\bibfnamefont
  {D.}~\bibnamefont {Tr\"{o}ndle}}, \bibinfo {author} {\bibfnamefont
  {F.}~\bibnamefont {Heine}}, \bibinfo {author} {\bibfnamefont
  {S.}~\bibnamefont {Seel}}, \bibinfo {author} {\bibfnamefont {P.}~\bibnamefont
  {Greulich}}, \bibinfo {author} {\bibfnamefont {H.}~\bibnamefont {Zech}},
  \bibinfo {author} {\bibfnamefont {B.}~\bibnamefont {G\"{u}tlich}}, \bibinfo
  {author} {\bibfnamefont {S.}~\bibnamefont {Philipp-May}}, \bibinfo {author}
  {\bibfnamefont {C.}~\bibnamefont {Marquardt}},\ and\ \bibinfo {author}
  {\bibfnamefont {G.}~\bibnamefont {Leuchs}},\ }\bibfield  {title} {\bibinfo
  {title} {Quantum-limited measurements of optical signals from a geostationary
  satellite},\ }\href {https://doi.org/10.1364/OPTICA.4.000611} {\bibfield
  {journal} {\bibinfo  {journal} {Optica}\ }\textbf {\bibinfo {volume} {4}},\
  \bibinfo {pages} {611} (\bibinfo {year} {2017})}\BibitemShut {NoStop}%
\bibitem [{\citenamefont {Yin}\ \emph {et~al.}(2013)\citenamefont {Yin},
  \citenamefont {Cao}, \citenamefont {Liu}, \citenamefont {Pan}, \citenamefont
  {Wang}, \citenamefont {Yang}, \citenamefont {Zhang}, \citenamefont {Yang},
  \citenamefont {Chen}, \citenamefont {Peng},\ and\ \citenamefont
  {Pan}}]{Yin:13}%
  \BibitemOpen
  \bibfield  {author} {\bibinfo {author} {\bibfnamefont {J.}~\bibnamefont
  {Yin}}, \bibinfo {author} {\bibfnamefont {Y.}~\bibnamefont {Cao}}, \bibinfo
  {author} {\bibfnamefont {S.-B.}\ \bibnamefont {Liu}}, \bibinfo {author}
  {\bibfnamefont {G.-S.}\ \bibnamefont {Pan}}, \bibinfo {author} {\bibfnamefont
  {J.-H.}\ \bibnamefont {Wang}}, \bibinfo {author} {\bibfnamefont
  {T.}~\bibnamefont {Yang}}, \bibinfo {author} {\bibfnamefont {Z.-P.}\
  \bibnamefont {Zhang}}, \bibinfo {author} {\bibfnamefont {F.-M.}\ \bibnamefont
  {Yang}}, \bibinfo {author} {\bibfnamefont {Y.-A.}\ \bibnamefont {Chen}},
  \bibinfo {author} {\bibfnamefont {C.-Z.}\ \bibnamefont {Peng}},\ and\
  \bibinfo {author} {\bibfnamefont {J.-W.}\ \bibnamefont {Pan}},\ }\bibfield
  {title} {\bibinfo {title} {Experimental quasi-single-photon transmission from
  satellite to earth},\ }\href {https://doi.org/10.1364/OE.21.020032}
  {\bibfield  {journal} {\bibinfo  {journal} {Opt. Express}\ }\textbf {\bibinfo
  {volume} {21}},\ \bibinfo {pages} {20032} (\bibinfo {year}
  {2013})}\BibitemShut {NoStop}%
\bibitem [{\citenamefont {Vallone}\ \emph {et~al.}(2015)\citenamefont
  {Vallone}, \citenamefont {Bacco}, \citenamefont {Dequal}, \citenamefont
  {Gaiarin}, \citenamefont {Luceri}, \citenamefont {Bianco},\ and\
  \citenamefont {Villoresi}}]{PhysRevLett.115.040502}%
  \BibitemOpen
  \bibfield  {author} {\bibinfo {author} {\bibfnamefont {G.}~\bibnamefont
  {Vallone}}, \bibinfo {author} {\bibfnamefont {D.}~\bibnamefont {Bacco}},
  \bibinfo {author} {\bibfnamefont {D.}~\bibnamefont {Dequal}}, \bibinfo
  {author} {\bibfnamefont {S.}~\bibnamefont {Gaiarin}}, \bibinfo {author}
  {\bibfnamefont {V.}~\bibnamefont {Luceri}}, \bibinfo {author} {\bibfnamefont
  {G.}~\bibnamefont {Bianco}},\ and\ \bibinfo {author} {\bibfnamefont
  {P.}~\bibnamefont {Villoresi}},\ }\bibfield  {title} {\bibinfo {title}
  {Experimental satellite quantum communications},\ }\href
  {https://doi.org/10.1103/PhysRevLett.115.040502} {\bibfield  {journal}
  {\bibinfo  {journal} {Phys. Rev. Lett.}\ }\textbf {\bibinfo {volume} {115}},\
  \bibinfo {pages} {040502} (\bibinfo {year} {2015})}\BibitemShut {NoStop}%
\bibitem [{\citenamefont {Vallone}\ \emph {et~al.}(2016)\citenamefont
  {Vallone}, \citenamefont {Dequal}, \citenamefont {Tomasin}, \citenamefont
  {Vedovato}, \citenamefont {Schiavon}, \citenamefont {Luceri}, \citenamefont
  {Bianco},\ and\ \citenamefont {Villoresi}}]{PhysRevLett.116.253601}%
  \BibitemOpen
  \bibfield  {author} {\bibinfo {author} {\bibfnamefont {G.}~\bibnamefont
  {Vallone}}, \bibinfo {author} {\bibfnamefont {D.}~\bibnamefont {Dequal}},
  \bibinfo {author} {\bibfnamefont {M.}~\bibnamefont {Tomasin}}, \bibinfo
  {author} {\bibfnamefont {F.}~\bibnamefont {Vedovato}}, \bibinfo {author}
  {\bibfnamefont {M.}~\bibnamefont {Schiavon}}, \bibinfo {author}
  {\bibfnamefont {V.}~\bibnamefont {Luceri}}, \bibinfo {author} {\bibfnamefont
  {G.}~\bibnamefont {Bianco}},\ and\ \bibinfo {author} {\bibfnamefont
  {P.}~\bibnamefont {Villoresi}},\ }\bibfield  {title} {\bibinfo {title}
  {Interference at the single photon level along satellite-ground channels},\
  }\href {https://doi.org/10.1103/PhysRevLett.116.253601} {\bibfield  {journal}
  {\bibinfo  {journal} {Phys. Rev. Lett.}\ }\textbf {\bibinfo {volume} {116}},\
  \bibinfo {pages} {253601} (\bibinfo {year} {2016})}\BibitemShut {NoStop}%
\bibitem [{\citenamefont {Yin}\ \emph {et~al.}(2017)\citenamefont {Yin},
  \citenamefont {Cao}, \citenamefont {Li}, \citenamefont {Liao}, \citenamefont
  {Zhang}, \citenamefont {Ren}, \citenamefont {Cai}, \citenamefont {Liu},
  \citenamefont {Li}, \citenamefont {Dai}, \citenamefont {Li}, \citenamefont
  {Lu}, \citenamefont {Gong}, \citenamefont {Xu}, \citenamefont {Li},
  \citenamefont {Li}, \citenamefont {Yin}, \citenamefont {Jiang}, \citenamefont
  {Li}, \citenamefont {Jia}, \citenamefont {Ren}, \citenamefont {He},
  \citenamefont {Zhou}, \citenamefont {Zhang}, \citenamefont {Wang},
  \citenamefont {Chang}, \citenamefont {Zhu}, \citenamefont {Liu},
  \citenamefont {Chen}, \citenamefont {Lu}, \citenamefont {Shu}, \citenamefont
  {Peng}, \citenamefont {Wang},\ and\ \citenamefont {Pan}}]{Yin1140}%
  \BibitemOpen
  \bibfield  {author} {\bibinfo {author} {\bibfnamefont {J.}~\bibnamefont
  {Yin}}, \bibinfo {author} {\bibfnamefont {Y.}~\bibnamefont {Cao}}, \bibinfo
  {author} {\bibfnamefont {Y.-H.}\ \bibnamefont {Li}}, \bibinfo {author}
  {\bibfnamefont {S.-K.}\ \bibnamefont {Liao}}, \bibinfo {author}
  {\bibfnamefont {L.}~\bibnamefont {Zhang}}, \bibinfo {author} {\bibfnamefont
  {J.-G.}\ \bibnamefont {Ren}}, \bibinfo {author} {\bibfnamefont {W.-Q.}\
  \bibnamefont {Cai}}, \bibinfo {author} {\bibfnamefont {W.-Y.}\ \bibnamefont
  {Liu}}, \bibinfo {author} {\bibfnamefont {B.}~\bibnamefont {Li}}, \bibinfo
  {author} {\bibfnamefont {H.}~\bibnamefont {Dai}}, \bibinfo {author}
  {\bibfnamefont {G.-B.}\ \bibnamefont {Li}}, \bibinfo {author} {\bibfnamefont
  {Q.-M.}\ \bibnamefont {Lu}}, \bibinfo {author} {\bibfnamefont {Y.-H.}\
  \bibnamefont {Gong}}, \bibinfo {author} {\bibfnamefont {Y.}~\bibnamefont
  {Xu}}, \bibinfo {author} {\bibfnamefont {S.-L.}\ \bibnamefont {Li}}, \bibinfo
  {author} {\bibfnamefont {F.-Z.}\ \bibnamefont {Li}}, \bibinfo {author}
  {\bibfnamefont {Y.-Y.}\ \bibnamefont {Yin}}, \bibinfo {author} {\bibfnamefont
  {Z.-Q.}\ \bibnamefont {Jiang}}, \bibinfo {author} {\bibfnamefont
  {M.}~\bibnamefont {Li}}, \bibinfo {author} {\bibfnamefont {J.-J.}\
  \bibnamefont {Jia}}, \bibinfo {author} {\bibfnamefont {G.}~\bibnamefont
  {Ren}}, \bibinfo {author} {\bibfnamefont {D.}~\bibnamefont {He}}, \bibinfo
  {author} {\bibfnamefont {Y.-L.}\ \bibnamefont {Zhou}}, \bibinfo {author}
  {\bibfnamefont {X.-X.}\ \bibnamefont {Zhang}}, \bibinfo {author}
  {\bibfnamefont {N.}~\bibnamefont {Wang}}, \bibinfo {author} {\bibfnamefont
  {X.}~\bibnamefont {Chang}}, \bibinfo {author} {\bibfnamefont {Z.-C.}\
  \bibnamefont {Zhu}}, \bibinfo {author} {\bibfnamefont {N.-L.}\ \bibnamefont
  {Liu}}, \bibinfo {author} {\bibfnamefont {Y.-A.}\ \bibnamefont {Chen}},
  \bibinfo {author} {\bibfnamefont {C.-Y.}\ \bibnamefont {Lu}}, \bibinfo
  {author} {\bibfnamefont {R.}~\bibnamefont {Shu}}, \bibinfo {author}
  {\bibfnamefont {C.-Z.}\ \bibnamefont {Peng}}, \bibinfo {author}
  {\bibfnamefont {J.-Y.}\ \bibnamefont {Wang}},\ and\ \bibinfo {author}
  {\bibfnamefont {J.-W.}\ \bibnamefont {Pan}},\ }\bibfield  {title} {\bibinfo
  {title} {Satellite-based entanglement distribution over 1200 kilometers},\
  }\href {https://doi.org/10.1126/science.aan3211} {\bibfield  {journal}
  {\bibinfo  {journal} {Science}\ }\textbf {\bibinfo {volume} {356}},\ \bibinfo
  {pages} {1140} (\bibinfo {year} {2017})}\BibitemShut {NoStop}%
\bibitem [{\citenamefont {Ren}\ \emph {et~al.}(2017)\citenamefont {Ren},
  \citenamefont {Xu}, \citenamefont {Yong}, \citenamefont {Zhang},
  \citenamefont {Liao}, \citenamefont {Yin}, \citenamefont {Liu}, \citenamefont
  {Cai}, \citenamefont {Yang}, \citenamefont {Li}, \citenamefont {Yang},
  \citenamefont {Han}, \citenamefont {Yao}, \citenamefont {Li}, \citenamefont
  {Wu}, \citenamefont {Wan}, \citenamefont {Liu}, \citenamefont {Liu},
  \citenamefont {Kuang}, \citenamefont {He}, \citenamefont {Shang},
  \citenamefont {Guo}, \citenamefont {Zheng}, \citenamefont {Tian},
  \citenamefont {Zhu}, \citenamefont {Liu}, \citenamefont {Lu}, \citenamefont
  {Shu}, \citenamefont {Chen}, \citenamefont {Peng}, \citenamefont {Wang},\
  and\ \citenamefont {Pan}}]{Ren2017}%
  \BibitemOpen
  \bibfield  {author} {\bibinfo {author} {\bibfnamefont {J.-G.}\ \bibnamefont
  {Ren}}, \bibinfo {author} {\bibfnamefont {P.}~\bibnamefont {Xu}}, \bibinfo
  {author} {\bibfnamefont {H.-L.}\ \bibnamefont {Yong}}, \bibinfo {author}
  {\bibfnamefont {L.}~\bibnamefont {Zhang}}, \bibinfo {author} {\bibfnamefont
  {S.-K.}\ \bibnamefont {Liao}}, \bibinfo {author} {\bibfnamefont
  {J.}~\bibnamefont {Yin}}, \bibinfo {author} {\bibfnamefont {W.-Y.}\
  \bibnamefont {Liu}}, \bibinfo {author} {\bibfnamefont {W.-Q.}\ \bibnamefont
  {Cai}}, \bibinfo {author} {\bibfnamefont {M.}~\bibnamefont {Yang}}, \bibinfo
  {author} {\bibfnamefont {L.}~\bibnamefont {Li}}, \bibinfo {author}
  {\bibfnamefont {K.-X.}\ \bibnamefont {Yang}}, \bibinfo {author}
  {\bibfnamefont {X.}~\bibnamefont {Han}}, \bibinfo {author} {\bibfnamefont
  {Y.-Q.}\ \bibnamefont {Yao}}, \bibinfo {author} {\bibfnamefont
  {J.}~\bibnamefont {Li}}, \bibinfo {author} {\bibfnamefont {H.-Y.}\
  \bibnamefont {Wu}}, \bibinfo {author} {\bibfnamefont {S.}~\bibnamefont
  {Wan}}, \bibinfo {author} {\bibfnamefont {L.}~\bibnamefont {Liu}}, \bibinfo
  {author} {\bibfnamefont {D.-Q.}\ \bibnamefont {Liu}}, \bibinfo {author}
  {\bibfnamefont {Y.-W.}\ \bibnamefont {Kuang}}, \bibinfo {author}
  {\bibfnamefont {Z.-P.}\ \bibnamefont {He}}, \bibinfo {author} {\bibfnamefont
  {P.}~\bibnamefont {Shang}}, \bibinfo {author} {\bibfnamefont
  {C.}~\bibnamefont {Guo}}, \bibinfo {author} {\bibfnamefont {R.-H.}\
  \bibnamefont {Zheng}}, \bibinfo {author} {\bibfnamefont {K.}~\bibnamefont
  {Tian}}, \bibinfo {author} {\bibfnamefont {Z.-C.}\ \bibnamefont {Zhu}},
  \bibinfo {author} {\bibfnamefont {N.-L.}\ \bibnamefont {Liu}}, \bibinfo
  {author} {\bibfnamefont {C.-Y.}\ \bibnamefont {Lu}}, \bibinfo {author}
  {\bibfnamefont {R.}~\bibnamefont {Shu}}, \bibinfo {author} {\bibfnamefont
  {Y.-A.}\ \bibnamefont {Chen}}, \bibinfo {author} {\bibfnamefont {C.-Z.}\
  \bibnamefont {Peng}}, \bibinfo {author} {\bibfnamefont {J.-Y.}\ \bibnamefont
  {Wang}},\ and\ \bibinfo {author} {\bibfnamefont {J.-W.}\ \bibnamefont
  {Pan}},\ }\bibfield  {title} {\bibinfo {title} {Ground-to-satellite quantum
  teleportation},\ }\href {https://doi.org/10.1038/nature23675} {\bibfield
  {journal} {\bibinfo  {journal} {Nature}\ }\textbf {\bibinfo {volume} {549}},\
  \bibinfo {pages} {70} (\bibinfo {year} {2017})}\BibitemShut {NoStop}%
\bibitem [{\citenamefont {Liao}\ \emph {et~al.}(2017)\citenamefont {Liao},
  \citenamefont {Cai}, \citenamefont {Liu}, \citenamefont {Zhang},
  \citenamefont {Li}, \citenamefont {Ren}, \citenamefont {Yin}, \citenamefont
  {Shen}, \citenamefont {Cao}, \citenamefont {Li}, \citenamefont {Li},
  \citenamefont {Chen}, \citenamefont {Sun}, \citenamefont {Jia}, \citenamefont
  {Wu}, \citenamefont {Jiang}, \citenamefont {Wang}, \citenamefont {Huang},
  \citenamefont {Wang}, \citenamefont {Zhou}, \citenamefont {Deng},
  \citenamefont {Xi}, \citenamefont {Ma}, \citenamefont {Hu}, \citenamefont
  {Zhang}, \citenamefont {Chen}, \citenamefont {Liu}, \citenamefont {Wang},
  \citenamefont {Zhu}, \citenamefont {Lu}, \citenamefont {Shu}, \citenamefont
  {Peng}, \citenamefont {Wang},\ and\ \citenamefont {Pan}}]{Liao2017}%
  \BibitemOpen
  \bibfield  {author} {\bibinfo {author} {\bibfnamefont {S.-K.}\ \bibnamefont
  {Liao}}, \bibinfo {author} {\bibfnamefont {W.-Q.}\ \bibnamefont {Cai}},
  \bibinfo {author} {\bibfnamefont {W.-Y.}\ \bibnamefont {Liu}}, \bibinfo
  {author} {\bibfnamefont {L.}~\bibnamefont {Zhang}}, \bibinfo {author}
  {\bibfnamefont {Y.}~\bibnamefont {Li}}, \bibinfo {author} {\bibfnamefont
  {J.-G.}\ \bibnamefont {Ren}}, \bibinfo {author} {\bibfnamefont
  {J.}~\bibnamefont {Yin}}, \bibinfo {author} {\bibfnamefont {Q.}~\bibnamefont
  {Shen}}, \bibinfo {author} {\bibfnamefont {Y.}~\bibnamefont {Cao}}, \bibinfo
  {author} {\bibfnamefont {Z.-P.}\ \bibnamefont {Li}}, \bibinfo {author}
  {\bibfnamefont {F.-Z.}\ \bibnamefont {Li}}, \bibinfo {author} {\bibfnamefont
  {X.-W.}\ \bibnamefont {Chen}}, \bibinfo {author} {\bibfnamefont {L.-H.}\
  \bibnamefont {Sun}}, \bibinfo {author} {\bibfnamefont {J.-J.}\ \bibnamefont
  {Jia}}, \bibinfo {author} {\bibfnamefont {J.-C.}\ \bibnamefont {Wu}},
  \bibinfo {author} {\bibfnamefont {X.-J.}\ \bibnamefont {Jiang}}, \bibinfo
  {author} {\bibfnamefont {J.-F.}\ \bibnamefont {Wang}}, \bibinfo {author}
  {\bibfnamefont {Y.-M.}\ \bibnamefont {Huang}}, \bibinfo {author}
  {\bibfnamefont {Q.}~\bibnamefont {Wang}}, \bibinfo {author} {\bibfnamefont
  {Y.-L.}\ \bibnamefont {Zhou}}, \bibinfo {author} {\bibfnamefont
  {L.}~\bibnamefont {Deng}}, \bibinfo {author} {\bibfnamefont {T.}~\bibnamefont
  {Xi}}, \bibinfo {author} {\bibfnamefont {L.}~\bibnamefont {Ma}}, \bibinfo
  {author} {\bibfnamefont {T.}~\bibnamefont {Hu}}, \bibinfo {author}
  {\bibfnamefont {Q.}~\bibnamefont {Zhang}}, \bibinfo {author} {\bibfnamefont
  {Y.-A.}\ \bibnamefont {Chen}}, \bibinfo {author} {\bibfnamefont {N.-L.}\
  \bibnamefont {Liu}}, \bibinfo {author} {\bibfnamefont {X.-B.}\ \bibnamefont
  {Wang}}, \bibinfo {author} {\bibfnamefont {Z.-C.}\ \bibnamefont {Zhu}},
  \bibinfo {author} {\bibfnamefont {C.-Y.}\ \bibnamefont {Lu}}, \bibinfo
  {author} {\bibfnamefont {R.}~\bibnamefont {Shu}}, \bibinfo {author}
  {\bibfnamefont {C.-Z.}\ \bibnamefont {Peng}}, \bibinfo {author}
  {\bibfnamefont {J.-Y.}\ \bibnamefont {Wang}},\ and\ \bibinfo {author}
  {\bibfnamefont {J.-W.}\ \bibnamefont {Pan}},\ }\bibfield  {title} {\bibinfo
  {title} {Satellite-to-ground quantum key distribution},\ }\href
  {https://doi.org/10.1038/nature23655} {\bibfield  {journal} {\bibinfo
  {journal} {Nature}\ }\textbf {\bibinfo {volume} {549}},\ \bibinfo {pages}
  {43} (\bibinfo {year} {2017})}\BibitemShut {NoStop}%
\bibitem [{\citenamefont {Liao}\ \emph {et~al.}(2018)\citenamefont {Liao},
  \citenamefont {Cai}, \citenamefont {Handsteiner}, \citenamefont {Liu},
  \citenamefont {Yin}, \citenamefont {Zhang}, \citenamefont {Rauch},
  \citenamefont {Fink}, \citenamefont {Ren}, \citenamefont {Liu}, \citenamefont
  {Li}, \citenamefont {Shen}, \citenamefont {Cao}, \citenamefont {Li},
  \citenamefont {Wang}, \citenamefont {Huang}, \citenamefont {Deng},
  \citenamefont {Xi}, \citenamefont {Ma}, \citenamefont {Hu}, \citenamefont
  {Li}, \citenamefont {Liu}, \citenamefont {Koidl}, \citenamefont {Wang},
  \citenamefont {Chen}, \citenamefont {Wang}, \citenamefont {Steindorfer},
  \citenamefont {Kirchner}, \citenamefont {Lu}, \citenamefont {Shu},
  \citenamefont {Ursin}, \citenamefont {Scheidl}, \citenamefont {Peng},
  \citenamefont {Wang}, \citenamefont {Zeilinger},\ and\ \citenamefont
  {Pan}}]{PhysRevLett.120.030501}%
  \BibitemOpen
  \bibfield  {author} {\bibinfo {author} {\bibfnamefont {S.-K.}\ \bibnamefont
  {Liao}}, \bibinfo {author} {\bibfnamefont {W.-Q.}\ \bibnamefont {Cai}},
  \bibinfo {author} {\bibfnamefont {J.}~\bibnamefont {Handsteiner}}, \bibinfo
  {author} {\bibfnamefont {B.}~\bibnamefont {Liu}}, \bibinfo {author}
  {\bibfnamefont {J.}~\bibnamefont {Yin}}, \bibinfo {author} {\bibfnamefont
  {L.}~\bibnamefont {Zhang}}, \bibinfo {author} {\bibfnamefont
  {D.}~\bibnamefont {Rauch}}, \bibinfo {author} {\bibfnamefont
  {M.}~\bibnamefont {Fink}}, \bibinfo {author} {\bibfnamefont {J.-G.}\
  \bibnamefont {Ren}}, \bibinfo {author} {\bibfnamefont {W.-Y.}\ \bibnamefont
  {Liu}}, \bibinfo {author} {\bibfnamefont {Y.}~\bibnamefont {Li}}, \bibinfo
  {author} {\bibfnamefont {Q.}~\bibnamefont {Shen}}, \bibinfo {author}
  {\bibfnamefont {Y.}~\bibnamefont {Cao}}, \bibinfo {author} {\bibfnamefont
  {F.-Z.}\ \bibnamefont {Li}}, \bibinfo {author} {\bibfnamefont {J.-F.}\
  \bibnamefont {Wang}}, \bibinfo {author} {\bibfnamefont {Y.-M.}\ \bibnamefont
  {Huang}}, \bibinfo {author} {\bibfnamefont {L.}~\bibnamefont {Deng}},
  \bibinfo {author} {\bibfnamefont {T.}~\bibnamefont {Xi}}, \bibinfo {author}
  {\bibfnamefont {L.}~\bibnamefont {Ma}}, \bibinfo {author} {\bibfnamefont
  {T.}~\bibnamefont {Hu}}, \bibinfo {author} {\bibfnamefont {L.}~\bibnamefont
  {Li}}, \bibinfo {author} {\bibfnamefont {N.-L.}\ \bibnamefont {Liu}},
  \bibinfo {author} {\bibfnamefont {F.}~\bibnamefont {Koidl}}, \bibinfo
  {author} {\bibfnamefont {P.}~\bibnamefont {Wang}}, \bibinfo {author}
  {\bibfnamefont {Y.-A.}\ \bibnamefont {Chen}}, \bibinfo {author}
  {\bibfnamefont {X.-B.}\ \bibnamefont {Wang}}, \bibinfo {author}
  {\bibfnamefont {M.}~\bibnamefont {Steindorfer}}, \bibinfo {author}
  {\bibfnamefont {G.}~\bibnamefont {Kirchner}}, \bibinfo {author}
  {\bibfnamefont {C.-Y.}\ \bibnamefont {Lu}}, \bibinfo {author} {\bibfnamefont
  {R.}~\bibnamefont {Shu}}, \bibinfo {author} {\bibfnamefont {R.}~\bibnamefont
  {Ursin}}, \bibinfo {author} {\bibfnamefont {T.}~\bibnamefont {Scheidl}},
  \bibinfo {author} {\bibfnamefont {C.-Z.}\ \bibnamefont {Peng}}, \bibinfo
  {author} {\bibfnamefont {J.-Y.}\ \bibnamefont {Wang}}, \bibinfo {author}
  {\bibfnamefont {A.}~\bibnamefont {Zeilinger}},\ and\ \bibinfo {author}
  {\bibfnamefont {J.-W.}\ \bibnamefont {Pan}},\ }\bibfield  {title} {\bibinfo
  {title} {Satellite-relayed intercontinental quantum network},\ }\href
  {https://doi.org/10.1103/PhysRevLett.120.030501} {\bibfield  {journal}
  {\bibinfo  {journal} {Phys. Rev. Lett.}\ }\textbf {\bibinfo {volume} {120}},\
  \bibinfo {pages} {030501} (\bibinfo {year} {2018})}\BibitemShut {NoStop}%
\bibitem [{\citenamefont {Armengol}\ \emph {et~al.}(2008)\citenamefont
  {Armengol}, \citenamefont {Furch}, \citenamefont {de~Matos}, \citenamefont
  {Minster}, \citenamefont {Cacciapuoti}, \citenamefont {Pfennigbauer},
  \citenamefont {Aspelmeyer}, \citenamefont {Jennewein}, \citenamefont {Ursin},
  \citenamefont {Schmitt-Manderbach}, \citenamefont {Baister}, \citenamefont
  {Rarity}, \citenamefont {Leeb}, \citenamefont {Barbieri}, \citenamefont
  {Weinfurter},\ and\ \citenamefont
  {Zeilinger}}]{10.1016/j.actaastro.2007.12.039}%
  \BibitemOpen
  \bibfield  {author} {\bibinfo {author} {\bibfnamefont {J.~M.~P.}\
  \bibnamefont {Armengol}}, \bibinfo {author} {\bibfnamefont {B.}~\bibnamefont
  {Furch}}, \bibinfo {author} {\bibfnamefont {C.~J.}\ \bibnamefont {de~Matos}},
  \bibinfo {author} {\bibfnamefont {O.}~\bibnamefont {Minster}}, \bibinfo
  {author} {\bibfnamefont {L.}~\bibnamefont {Cacciapuoti}}, \bibinfo {author}
  {\bibfnamefont {M.}~\bibnamefont {Pfennigbauer}}, \bibinfo {author}
  {\bibfnamefont {M.}~\bibnamefont {Aspelmeyer}}, \bibinfo {author}
  {\bibfnamefont {T.}~\bibnamefont {Jennewein}}, \bibinfo {author}
  {\bibfnamefont {R.}~\bibnamefont {Ursin}}, \bibinfo {author} {\bibfnamefont
  {T.}~\bibnamefont {Schmitt-Manderbach}}, \bibinfo {author} {\bibfnamefont
  {G.}~\bibnamefont {Baister}}, \bibinfo {author} {\bibfnamefont
  {J.}~\bibnamefont {Rarity}}, \bibinfo {author} {\bibfnamefont
  {W.}~\bibnamefont {Leeb}}, \bibinfo {author} {\bibfnamefont {C.}~\bibnamefont
  {Barbieri}}, \bibinfo {author} {\bibfnamefont {H.}~\bibnamefont
  {Weinfurter}},\ and\ \bibinfo {author} {\bibfnamefont {A.}~\bibnamefont
  {Zeilinger}},\ }\bibfield  {title} {\bibinfo {title} {{Quantum communications
  at ESA: Towards a space experiment on the ISS}},\ }\href
  {https://doi.org/10.1016/j.actaastro.2007.12.039} {\bibfield  {journal}
  {\bibinfo  {journal} {Acta Astronautica}\ }\textbf {\bibinfo {volume} {63}},\
  \bibinfo {pages} {165} (\bibinfo {year} {2008})}\BibitemShut {NoStop}%
\bibitem [{\citenamefont {Jennewein}\ \emph
  {et~al.}(2014{\natexlab{a}})\citenamefont {Jennewein}, \citenamefont
  {Bourgoin}, \citenamefont {Higgins}, \citenamefont {Holloway}, \citenamefont
  {Meyer-Scott}, \citenamefont {Erven}, \citenamefont {Heim}, \citenamefont
  {Yan}, \citenamefont {Hübel}, \citenamefont {Weihs}, \citenamefont {Choi},
  \citenamefont {D'Souza}, \citenamefont {Hudson},\ and\ \citenamefont
  {Laflamme}}]{10.1117/12.2041693}%
  \BibitemOpen
  \bibfield  {author} {\bibinfo {author} {\bibfnamefont {T.}~\bibnamefont
  {Jennewein}}, \bibinfo {author} {\bibfnamefont {J.~P.}\ \bibnamefont
  {Bourgoin}}, \bibinfo {author} {\bibfnamefont {B.}~\bibnamefont {Higgins}},
  \bibinfo {author} {\bibfnamefont {C.}~\bibnamefont {Holloway}}, \bibinfo
  {author} {\bibfnamefont {E.}~\bibnamefont {Meyer-Scott}}, \bibinfo {author}
  {\bibfnamefont {C.}~\bibnamefont {Erven}}, \bibinfo {author} {\bibfnamefont
  {B.}~\bibnamefont {Heim}}, \bibinfo {author} {\bibfnamefont {Z.}~\bibnamefont
  {Yan}}, \bibinfo {author} {\bibfnamefont {H.}~\bibnamefont {Hübel}},
  \bibinfo {author} {\bibfnamefont {G.}~\bibnamefont {Weihs}}, \bibinfo
  {author} {\bibfnamefont {E.}~\bibnamefont {Choi}}, \bibinfo {author}
  {\bibfnamefont {I.}~\bibnamefont {D'Souza}}, \bibinfo {author} {\bibfnamefont
  {D.}~\bibnamefont {Hudson}},\ and\ \bibinfo {author} {\bibfnamefont
  {R.}~\bibnamefont {Laflamme}},\ }\bibfield  {title} {\bibinfo {title}
  {{QEYSSAT: a mission proposal for a quantum receiver in space}},\ }in\ \href
  {https://doi.org/10.1117/12.2041693} {\emph {\bibinfo {booktitle} {Advances
  in Photonics of Quantum Computing, Memory, and Communication VII}}},\ Vol.\
  \bibinfo {volume} {8997},\ \bibinfo {editor} {edited by\ \bibinfo {editor}
  {\bibfnamefont {Z.~U.}\ \bibnamefont {Hasan}}, \bibinfo {editor}
  {\bibfnamefont {P.~R.}\ \bibnamefont {Hemmer}}, \bibinfo {editor}
  {\bibfnamefont {H.}~\bibnamefont {Lee}},\ and\ \bibinfo {editor}
  {\bibfnamefont {C.~M.}\ \bibnamefont {Santori}}},\ \bibinfo {organization}
  {International Society for Optics and Photonics}\ (\bibinfo  {publisher}
  {SPIE},\ \bibinfo {year} {2014})\ pp.\ \bibinfo {pages} {21 --
  27}\BibitemShut {NoStop}%
\bibitem [{\citenamefont {Jennewein}\ \emph
  {et~al.}(2014{\natexlab{b}})\citenamefont {Jennewein}, \citenamefont {Grant},
  \citenamefont {Choi}, \citenamefont {Pugh}, \citenamefont {Holloway},
  \citenamefont {Bourgoin}, \citenamefont {Hakima}, \citenamefont {Higgins},\
  and\ \citenamefont {Zee}}]{10.1117/12.2067548}%
  \BibitemOpen
  \bibfield  {author} {\bibinfo {author} {\bibfnamefont {T.}~\bibnamefont
  {Jennewein}}, \bibinfo {author} {\bibfnamefont {C.}~\bibnamefont {Grant}},
  \bibinfo {author} {\bibfnamefont {E.}~\bibnamefont {Choi}}, \bibinfo {author}
  {\bibfnamefont {C.}~\bibnamefont {Pugh}}, \bibinfo {author} {\bibfnamefont
  {C.}~\bibnamefont {Holloway}}, \bibinfo {author} {\bibfnamefont
  {J.}~\bibnamefont {Bourgoin}}, \bibinfo {author} {\bibfnamefont
  {H.}~\bibnamefont {Hakima}}, \bibinfo {author} {\bibfnamefont
  {B.}~\bibnamefont {Higgins}},\ and\ \bibinfo {author} {\bibfnamefont
  {R.}~\bibnamefont {Zee}},\ }\bibfield  {title} {\bibinfo {title} {{The
  NanoQEY mission: ground to space quantum key and entanglement distribution
  using a nanosatellite}},\ }in\ \href {https://doi.org/10.1117/12.2067548}
  {\emph {\bibinfo {booktitle} {Emerging Technologies in Security and Defence
  II; and Quantum-Physics-based Information Security III}}},\ Vol.\ \bibinfo
  {volume} {9254},\ \bibinfo {editor} {edited by\ \bibinfo {editor}
  {\bibfnamefont {K.~L.}\ \bibnamefont {Lewis}}, \bibinfo {editor}
  {\bibfnamefont {R.~C.}\ \bibnamefont {Hollins}}, \bibinfo {editor}
  {\bibfnamefont {T.~J.}\ \bibnamefont {Merlet}}, \bibinfo {editor}
  {\bibfnamefont {A.}~\bibnamefont {Toet}}, \bibinfo {editor} {\bibfnamefont
  {M.~T.}\ \bibnamefont {Gruneisen}}, \bibinfo {editor} {\bibfnamefont
  {M.}~\bibnamefont {Dusek}},\ and\ \bibinfo {editor} {\bibfnamefont {J.~G.}\
  \bibnamefont {Rarity}}},\ \bibinfo {organization} {International Society for
  Optics and Photonics}\ (\bibinfo  {publisher} {SPIE},\ \bibinfo {year}
  {2014})\ pp.\ \bibinfo {pages} {1 -- 6}\BibitemShut {NoStop}%
\bibitem [{\citenamefont {Oi}\ \emph {et~al.}(2017)\citenamefont {Oi},
  \citenamefont {Ling}, \citenamefont {Vallone}, \citenamefont {Villoresi},
  \citenamefont {Greenland}, \citenamefont {Kerr}, \citenamefont {Macdonald},
  \citenamefont {Weinfurter}, \citenamefont {Kuiper}, \citenamefont {Charbon},\
  and\ \citenamefont {Ursin}}]{Oi2017}%
  \BibitemOpen
  \bibfield  {author} {\bibinfo {author} {\bibfnamefont {D.~K.}\ \bibnamefont
  {Oi}}, \bibinfo {author} {\bibfnamefont {A.}~\bibnamefont {Ling}}, \bibinfo
  {author} {\bibfnamefont {G.}~\bibnamefont {Vallone}}, \bibinfo {author}
  {\bibfnamefont {P.}~\bibnamefont {Villoresi}}, \bibinfo {author}
  {\bibfnamefont {S.}~\bibnamefont {Greenland}}, \bibinfo {author}
  {\bibfnamefont {E.}~\bibnamefont {Kerr}}, \bibinfo {author} {\bibfnamefont
  {M.}~\bibnamefont {Macdonald}}, \bibinfo {author} {\bibfnamefont
  {H.}~\bibnamefont {Weinfurter}}, \bibinfo {author} {\bibfnamefont
  {H.}~\bibnamefont {Kuiper}}, \bibinfo {author} {\bibfnamefont
  {E.}~\bibnamefont {Charbon}},\ and\ \bibinfo {author} {\bibfnamefont
  {R.}~\bibnamefont {Ursin}},\ }\bibfield  {title} {\bibinfo {title} {{CubeSat
  quantum communications mission}},\ }\href
  {https://doi.org/10.1140/epjqt/s40507-017-0060-1} {\bibfield  {journal}
  {\bibinfo  {journal} {EPJ Quantum Technology}\ }\textbf {\bibinfo {volume}
  {4}},\ \bibinfo {pages} {6} (\bibinfo {year} {2017})}\BibitemShut {NoStop}%
\bibitem [{\citenamefont {Kerstel}\ \emph {et~al.}(2018)\citenamefont
  {Kerstel}, \citenamefont {Gardelein}, \citenamefont {Barthelemy},
  \citenamefont {Fink}, \citenamefont {Joshi},\ and\ \citenamefont
  {Ursin}}]{Kerstel2018}%
  \BibitemOpen
  \bibfield  {author} {\bibinfo {author} {\bibfnamefont {E.}~\bibnamefont
  {Kerstel}}, \bibinfo {author} {\bibfnamefont {A.}~\bibnamefont {Gardelein}},
  \bibinfo {author} {\bibfnamefont {M.}~\bibnamefont {Barthelemy}}, \bibinfo
  {author} {\bibfnamefont {M.}~\bibnamefont {Fink}}, \bibinfo {author}
  {\bibfnamefont {S.~K.}\ \bibnamefont {Joshi}},\ and\ \bibinfo {author}
  {\bibfnamefont {R.}~\bibnamefont {Ursin}},\ }\bibfield  {title} {\bibinfo
  {title} {{Nanobob: a CubeSat mission concept for quantum communication
  experiments in an uplink configuration}},\ }\href
  {https://doi.org/10.1140/epjqt/s40507-018-0070-7} {\bibfield  {journal}
  {\bibinfo  {journal} {EPJ Quantum Technology}\ }\textbf {\bibinfo {volume}
  {5}},\ \bibinfo {pages} {6} (\bibinfo {year} {2018})}\BibitemShut {NoStop}%
\bibitem [{\citenamefont {Haber}\ \emph {et~al.}(2018)\citenamefont {Haber},
  \citenamefont {Garbe}, \citenamefont {Busch}, \citenamefont {Rosenfeld},\
  and\ \citenamefont {Schilling}}]{Haber2018QubeA}%
  \BibitemOpen
  \bibfield  {author} {\bibinfo {author} {\bibfnamefont {R.}~\bibnamefont
  {Haber}}, \bibinfo {author} {\bibfnamefont {D.}~\bibnamefont {Garbe}},
  \bibinfo {author} {\bibfnamefont {S.}~\bibnamefont {Busch}}, \bibinfo
  {author} {\bibfnamefont {W.}~\bibnamefont {Rosenfeld}},\ and\ \bibinfo
  {author} {\bibfnamefont {K.}~\bibnamefont {Schilling}},\ }\bibfield  {title}
  {\bibinfo {title} {{Qube - A CubeSat for Quantum Key Distribution
  Experiments}},\ }in\ \href@noop {} {\emph {\bibinfo {booktitle} {Proceedings
  of the AIAA/USU Conference on Small Satellites}}},\ Vol.~\bibinfo {volume}
  {49}\ (\bibinfo {year} {2018})\BibitemShut {NoStop}%
\bibitem [{\citenamefont {Sansoni}\ \emph {et~al.}(2017)\citenamefont
  {Sansoni}, \citenamefont {Luo}, \citenamefont {Eigner}, \citenamefont
  {Ricken}, \citenamefont {Quiring}, \citenamefont {Herrmann},\ and\
  \citenamefont {Silberhorn}}]{sansoni2017two}%
  \BibitemOpen
  \bibfield  {author} {\bibinfo {author} {\bibfnamefont {L.}~\bibnamefont
  {Sansoni}}, \bibinfo {author} {\bibfnamefont {K.~H.}\ \bibnamefont {Luo}},
  \bibinfo {author} {\bibfnamefont {C.}~\bibnamefont {Eigner}}, \bibinfo
  {author} {\bibfnamefont {R.}~\bibnamefont {Ricken}}, \bibinfo {author}
  {\bibfnamefont {V.}~\bibnamefont {Quiring}}, \bibinfo {author} {\bibfnamefont
  {H.}~\bibnamefont {Herrmann}},\ and\ \bibinfo {author} {\bibfnamefont
  {C.}~\bibnamefont {Silberhorn}},\ }\bibfield  {title} {\bibinfo {title} {A
  two-channel, spectrally degenerate polarization entangled source on chip},\
  }\href@noop {} {\bibfield  {journal} {\bibinfo  {journal} {npj Quantum
  Information}\ }\textbf {\bibinfo {volume} {3}},\ \bibinfo {pages} {1}
  (\bibinfo {year} {2017})}\BibitemShut {NoStop}%
\bibitem [{\citenamefont {Orieux}\ \emph {et~al.}(2013)\citenamefont {Orieux},
  \citenamefont {Eckstein}, \citenamefont {Lema{\^\i}tre}, \citenamefont
  {Filloux}, \citenamefont {Favero}, \citenamefont {Leo}, \citenamefont
  {Coudreau}, \citenamefont {Keller}, \citenamefont {Milman},\ and\
  \citenamefont {Ducci}}]{orieux2013direct}%
  \BibitemOpen
  \bibfield  {author} {\bibinfo {author} {\bibfnamefont {A.}~\bibnamefont
  {Orieux}}, \bibinfo {author} {\bibfnamefont {A.}~\bibnamefont {Eckstein}},
  \bibinfo {author} {\bibfnamefont {A.}~\bibnamefont {Lema{\^\i}tre}}, \bibinfo
  {author} {\bibfnamefont {P.}~\bibnamefont {Filloux}}, \bibinfo {author}
  {\bibfnamefont {I.}~\bibnamefont {Favero}}, \bibinfo {author} {\bibfnamefont
  {G.}~\bibnamefont {Leo}}, \bibinfo {author} {\bibfnamefont {T.}~\bibnamefont
  {Coudreau}}, \bibinfo {author} {\bibfnamefont {A.}~\bibnamefont {Keller}},
  \bibinfo {author} {\bibfnamefont {P.}~\bibnamefont {Milman}},\ and\ \bibinfo
  {author} {\bibfnamefont {S.}~\bibnamefont {Ducci}},\ }\bibfield  {title}
  {\bibinfo {title} {Direct bell states generation on a iii-v semiconductor
  chip at room temperature},\ }\href@noop {} {\bibfield  {journal} {\bibinfo
  {journal} {Physical review letters}\ }\textbf {\bibinfo {volume} {110}},\
  \bibinfo {pages} {160502} (\bibinfo {year} {2013})}\BibitemShut {NoStop}%
\bibitem [{\citenamefont {Qiang}\ \emph {et~al.}(2018)\citenamefont {Qiang},
  \citenamefont {Zhou}, \citenamefont {Wang}, \citenamefont {Wilkes},
  \citenamefont {Loke}, \citenamefont {O’Gara}, \citenamefont {Kling},
  \citenamefont {Marshall}, \citenamefont {Santagati}, \citenamefont {Ralph}
  \emph {et~al.}}]{qiang2018large}%
  \BibitemOpen
  \bibfield  {author} {\bibinfo {author} {\bibfnamefont {X.}~\bibnamefont
  {Qiang}}, \bibinfo {author} {\bibfnamefont {X.}~\bibnamefont {Zhou}},
  \bibinfo {author} {\bibfnamefont {J.}~\bibnamefont {Wang}}, \bibinfo {author}
  {\bibfnamefont {C.~M.}\ \bibnamefont {Wilkes}}, \bibinfo {author}
  {\bibfnamefont {T.}~\bibnamefont {Loke}}, \bibinfo {author} {\bibfnamefont
  {S.}~\bibnamefont {O’Gara}}, \bibinfo {author} {\bibfnamefont
  {L.}~\bibnamefont {Kling}}, \bibinfo {author} {\bibfnamefont {G.~D.}\
  \bibnamefont {Marshall}}, \bibinfo {author} {\bibfnamefont {R.}~\bibnamefont
  {Santagati}}, \bibinfo {author} {\bibfnamefont {T.~C.}\ \bibnamefont
  {Ralph}}, \emph {et~al.},\ }\bibfield  {title} {\bibinfo {title} {Large-scale
  silicon quantum photonics implementing arbitrary two-qubit processing},\
  }\href@noop {} {\bibfield  {journal} {\bibinfo  {journal} {Nature photonics}\
  }\textbf {\bibinfo {volume} {12}},\ \bibinfo {pages} {534} (\bibinfo {year}
  {2018})}\BibitemShut {NoStop}%
\bibitem [{\citenamefont {Carolan}\ \emph {et~al.}(2015)\citenamefont
  {Carolan}, \citenamefont {Harrold}, \citenamefont {Sparrow}, \citenamefont
  {Mart{\'\i}n-L{\'o}pez}, \citenamefont {Russell}, \citenamefont
  {Silverstone}, \citenamefont {Shadbolt}, \citenamefont {Matsuda},
  \citenamefont {Oguma}, \citenamefont {Itoh} \emph
  {et~al.}}]{carolan2015universal}%
  \BibitemOpen
  \bibfield  {author} {\bibinfo {author} {\bibfnamefont {J.}~\bibnamefont
  {Carolan}}, \bibinfo {author} {\bibfnamefont {C.}~\bibnamefont {Harrold}},
  \bibinfo {author} {\bibfnamefont {C.}~\bibnamefont {Sparrow}}, \bibinfo
  {author} {\bibfnamefont {E.}~\bibnamefont {Mart{\'\i}n-L{\'o}pez}}, \bibinfo
  {author} {\bibfnamefont {N.~J.}\ \bibnamefont {Russell}}, \bibinfo {author}
  {\bibfnamefont {J.~W.}\ \bibnamefont {Silverstone}}, \bibinfo {author}
  {\bibfnamefont {P.~J.}\ \bibnamefont {Shadbolt}}, \bibinfo {author}
  {\bibfnamefont {N.}~\bibnamefont {Matsuda}}, \bibinfo {author} {\bibfnamefont
  {M.}~\bibnamefont {Oguma}}, \bibinfo {author} {\bibfnamefont
  {M.}~\bibnamefont {Itoh}}, \emph {et~al.},\ }\bibfield  {title} {\bibinfo
  {title} {Universal linear optics},\ }\href@noop {} {\bibfield  {journal}
  {\bibinfo  {journal} {Science}\ }\textbf {\bibinfo {volume} {349}},\ \bibinfo
  {pages} {711} (\bibinfo {year} {2015})}\BibitemShut {NoStop}%
\bibitem [{\citenamefont {Najafi}\ \emph {et~al.}(2015)\citenamefont {Najafi},
  \citenamefont {Mower}, \citenamefont {Harris}, \citenamefont {Bellei},
  \citenamefont {Dane}, \citenamefont {Lee}, \citenamefont {Hu}, \citenamefont
  {Kharel}, \citenamefont {Marsili}, \citenamefont {Assefa} \emph
  {et~al.}}]{najafi2015chip}%
  \BibitemOpen
  \bibfield  {author} {\bibinfo {author} {\bibfnamefont {F.}~\bibnamefont
  {Najafi}}, \bibinfo {author} {\bibfnamefont {J.}~\bibnamefont {Mower}},
  \bibinfo {author} {\bibfnamefont {N.~C.}\ \bibnamefont {Harris}}, \bibinfo
  {author} {\bibfnamefont {F.}~\bibnamefont {Bellei}}, \bibinfo {author}
  {\bibfnamefont {A.}~\bibnamefont {Dane}}, \bibinfo {author} {\bibfnamefont
  {C.}~\bibnamefont {Lee}}, \bibinfo {author} {\bibfnamefont {X.}~\bibnamefont
  {Hu}}, \bibinfo {author} {\bibfnamefont {P.}~\bibnamefont {Kharel}}, \bibinfo
  {author} {\bibfnamefont {F.}~\bibnamefont {Marsili}}, \bibinfo {author}
  {\bibfnamefont {S.}~\bibnamefont {Assefa}}, \emph {et~al.},\ }\bibfield
  {title} {\bibinfo {title} {On-chip detection of non-classical light by
  scalable integration of single-photon detectors},\ }\href@noop {} {\bibfield
  {journal} {\bibinfo  {journal} {Nature communications}\ }\textbf {\bibinfo
  {volume} {6}},\ \bibinfo {pages} {1} (\bibinfo {year} {2015})}\BibitemShut
  {NoStop}%
\bibitem [{\citenamefont {Marshall}\ \emph {et~al.}(2009)\citenamefont
  {Marshall}, \citenamefont {Politi}, \citenamefont {Matthews}, \citenamefont
  {Dekker}, \citenamefont {Ams}, \citenamefont {Withford},\ and\ \citenamefont
  {O’Brien}}]{marshall2009laser}%
  \BibitemOpen
  \bibfield  {author} {\bibinfo {author} {\bibfnamefont {G.~D.}\ \bibnamefont
  {Marshall}}, \bibinfo {author} {\bibfnamefont {A.}~\bibnamefont {Politi}},
  \bibinfo {author} {\bibfnamefont {J.~C.}\ \bibnamefont {Matthews}}, \bibinfo
  {author} {\bibfnamefont {P.}~\bibnamefont {Dekker}}, \bibinfo {author}
  {\bibfnamefont {M.}~\bibnamefont {Ams}}, \bibinfo {author} {\bibfnamefont
  {M.~J.}\ \bibnamefont {Withford}},\ and\ \bibinfo {author} {\bibfnamefont
  {J.~L.}\ \bibnamefont {O’Brien}},\ }\bibfield  {title} {\bibinfo {title}
  {Laser written waveguide photonic quantum circuits},\ }\href@noop {}
  {\bibfield  {journal} {\bibinfo  {journal} {Optics express}\ }\textbf
  {\bibinfo {volume} {17}},\ \bibinfo {pages} {12546} (\bibinfo {year}
  {2009})}\BibitemShut {NoStop}%
\bibitem [{\citenamefont {Gr{\"a}fe}\ \emph {et~al.}(2014)\citenamefont
  {Gr{\"a}fe}, \citenamefont {Heilmann}, \citenamefont {Perez-Leija},
  \citenamefont {Keil}, \citenamefont {Dreisow}, \citenamefont {Heinrich},
  \citenamefont {Moya-Cessa}, \citenamefont {Nolte}, \citenamefont
  {Christodoulides},\ and\ \citenamefont {Szameit}}]{grafe2014chip}%
  \BibitemOpen
  \bibfield  {author} {\bibinfo {author} {\bibfnamefont {M.}~\bibnamefont
  {Gr{\"a}fe}}, \bibinfo {author} {\bibfnamefont {R.}~\bibnamefont {Heilmann}},
  \bibinfo {author} {\bibfnamefont {A.}~\bibnamefont {Perez-Leija}}, \bibinfo
  {author} {\bibfnamefont {R.}~\bibnamefont {Keil}}, \bibinfo {author}
  {\bibfnamefont {F.}~\bibnamefont {Dreisow}}, \bibinfo {author} {\bibfnamefont
  {M.}~\bibnamefont {Heinrich}}, \bibinfo {author} {\bibfnamefont
  {H.}~\bibnamefont {Moya-Cessa}}, \bibinfo {author} {\bibfnamefont
  {S.}~\bibnamefont {Nolte}}, \bibinfo {author} {\bibfnamefont {D.~N.}\
  \bibnamefont {Christodoulides}},\ and\ \bibinfo {author} {\bibfnamefont
  {A.}~\bibnamefont {Szameit}},\ }\bibfield  {title} {\bibinfo {title} {On-chip
  generation of high-order single-photon w-states},\ }\href@noop {} {\bibfield
  {journal} {\bibinfo  {journal} {Nature Photonics}\ }\textbf {\bibinfo
  {volume} {8}},\ \bibinfo {pages} {791} (\bibinfo {year} {2014})}\BibitemShut
  {NoStop}%
\bibitem [{\citenamefont {Atzeni}\ \emph {et~al.}(2018)\citenamefont {Atzeni},
  \citenamefont {Rab}, \citenamefont {Corrielli}, \citenamefont {Polino},
  \citenamefont {Valeri}, \citenamefont {Mataloni}, \citenamefont {Spagnolo},
  \citenamefont {Crespi}, \citenamefont {Sciarrino},\ and\ \citenamefont
  {Osellame}}]{atzeni2018integrated}%
  \BibitemOpen
  \bibfield  {author} {\bibinfo {author} {\bibfnamefont {S.}~\bibnamefont
  {Atzeni}}, \bibinfo {author} {\bibfnamefont {A.~S.}\ \bibnamefont {Rab}},
  \bibinfo {author} {\bibfnamefont {G.}~\bibnamefont {Corrielli}}, \bibinfo
  {author} {\bibfnamefont {E.}~\bibnamefont {Polino}}, \bibinfo {author}
  {\bibfnamefont {M.}~\bibnamefont {Valeri}}, \bibinfo {author} {\bibfnamefont
  {P.}~\bibnamefont {Mataloni}}, \bibinfo {author} {\bibfnamefont
  {N.}~\bibnamefont {Spagnolo}}, \bibinfo {author} {\bibfnamefont
  {A.}~\bibnamefont {Crespi}}, \bibinfo {author} {\bibfnamefont
  {F.}~\bibnamefont {Sciarrino}},\ and\ \bibinfo {author} {\bibfnamefont
  {R.}~\bibnamefont {Osellame}},\ }\bibfield  {title} {\bibinfo {title}
  {Integrated sources of entangled photons at the telecom wavelength in
  femtosecond-laser-written circuits},\ }\href@noop {} {\bibfield  {journal}
  {\bibinfo  {journal} {Optica}\ }\textbf {\bibinfo {volume} {5}},\ \bibinfo
  {pages} {311} (\bibinfo {year} {2018})}\BibitemShut {NoStop}%
\bibitem [{\citenamefont {Polino}\ \emph {et~al.}(2019)\citenamefont {Polino},
  \citenamefont {Riva}, \citenamefont {Valeri}, \citenamefont {Silvestri},
  \citenamefont {Corrielli}, \citenamefont {Crespi}, \citenamefont {Spagnolo},
  \citenamefont {Osellame},\ and\ \citenamefont
  {Sciarrino}}]{polino2019experimental}%
  \BibitemOpen
  \bibfield  {author} {\bibinfo {author} {\bibfnamefont {E.}~\bibnamefont
  {Polino}}, \bibinfo {author} {\bibfnamefont {M.}~\bibnamefont {Riva}},
  \bibinfo {author} {\bibfnamefont {M.}~\bibnamefont {Valeri}}, \bibinfo
  {author} {\bibfnamefont {R.}~\bibnamefont {Silvestri}}, \bibinfo {author}
  {\bibfnamefont {G.}~\bibnamefont {Corrielli}}, \bibinfo {author}
  {\bibfnamefont {A.}~\bibnamefont {Crespi}}, \bibinfo {author} {\bibfnamefont
  {N.}~\bibnamefont {Spagnolo}}, \bibinfo {author} {\bibfnamefont
  {R.}~\bibnamefont {Osellame}},\ and\ \bibinfo {author} {\bibfnamefont
  {F.}~\bibnamefont {Sciarrino}},\ }\bibfield  {title} {\bibinfo {title}
  {Experimental multiphase estimation on a chip},\ }\href@noop {} {\bibfield
  {journal} {\bibinfo  {journal} {Optica}\ }\textbf {\bibinfo {volume} {6}},\
  \bibinfo {pages} {288} (\bibinfo {year} {2019})}\BibitemShut {NoStop}%
\bibitem [{\citenamefont {Ant{\'o}n}\ \emph {et~al.}(2019)\citenamefont
  {Ant{\'o}n}, \citenamefont {Loredo}, \citenamefont {Coppola}, \citenamefont
  {Ollivier}, \citenamefont {Viggianiello}, \citenamefont {Harouri},
  \citenamefont {Somaschi}, \citenamefont {Crespi}, \citenamefont {Sagnes},
  \citenamefont {Lemaitre} \emph {et~al.}}]{anton2019interfacing}%
  \BibitemOpen
  \bibfield  {author} {\bibinfo {author} {\bibfnamefont {C.}~\bibnamefont
  {Ant{\'o}n}}, \bibinfo {author} {\bibfnamefont {J.~C.}\ \bibnamefont
  {Loredo}}, \bibinfo {author} {\bibfnamefont {G.}~\bibnamefont {Coppola}},
  \bibinfo {author} {\bibfnamefont {H.}~\bibnamefont {Ollivier}}, \bibinfo
  {author} {\bibfnamefont {N.}~\bibnamefont {Viggianiello}}, \bibinfo {author}
  {\bibfnamefont {A.}~\bibnamefont {Harouri}}, \bibinfo {author} {\bibfnamefont
  {N.}~\bibnamefont {Somaschi}}, \bibinfo {author} {\bibfnamefont
  {A.}~\bibnamefont {Crespi}}, \bibinfo {author} {\bibfnamefont
  {I.}~\bibnamefont {Sagnes}}, \bibinfo {author} {\bibfnamefont
  {A.}~\bibnamefont {Lemaitre}}, \emph {et~al.},\ }\bibfield  {title} {\bibinfo
  {title} {Interfacing scalable photonic platforms: Solid-state based
  multi-photon interference in a reconfigurable glass chip},\ }\href@noop {}
  {\bibfield  {journal} {\bibinfo  {journal} {Optica}\ }\textbf {\bibinfo
  {volume} {6}},\ \bibinfo {pages} {1471} (\bibinfo {year} {2019})}\BibitemShut
  {NoStop}%
\bibitem [{\citenamefont {Della~Valle}\ \emph {et~al.}(2008)\citenamefont
  {Della~Valle}, \citenamefont {Osellame},\ and\ \citenamefont
  {Laporta}}]{della2008micromachining}%
  \BibitemOpen
  \bibfield  {author} {\bibinfo {author} {\bibfnamefont {G.}~\bibnamefont
  {Della~Valle}}, \bibinfo {author} {\bibfnamefont {R.}~\bibnamefont
  {Osellame}},\ and\ \bibinfo {author} {\bibfnamefont {P.}~\bibnamefont
  {Laporta}},\ }\bibfield  {title} {\bibinfo {title} {Micromachining of
  photonic devices by femtosecond laser pulses},\ }\href@noop {} {\bibfield
  {journal} {\bibinfo  {journal} {Journal of Optics A: Pure and Applied
  Optics}\ }\textbf {\bibinfo {volume} {11}},\ \bibinfo {pages} {013001}
  (\bibinfo {year} {2008})}\BibitemShut {NoStop}%
\bibitem [{\citenamefont {Corrielli}\ \emph {et~al.}(2018)\citenamefont
  {Corrielli}, \citenamefont {Atzeni}, \citenamefont {Piacentini},
  \citenamefont {Pitsios}, \citenamefont {Crespi},\ and\ \citenamefont
  {Osellame}}]{corrielli2018}%
  \BibitemOpen
  \bibfield  {author} {\bibinfo {author} {\bibfnamefont {G.}~\bibnamefont
  {Corrielli}}, \bibinfo {author} {\bibfnamefont {S.}~\bibnamefont {Atzeni}},
  \bibinfo {author} {\bibfnamefont {S.}~\bibnamefont {Piacentini}}, \bibinfo
  {author} {\bibfnamefont {I.}~\bibnamefont {Pitsios}}, \bibinfo {author}
  {\bibfnamefont {A.}~\bibnamefont {Crespi}},\ and\ \bibinfo {author}
  {\bibfnamefont {R.}~\bibnamefont {Osellame}},\ }\bibfield  {title} {\bibinfo
  {title} {Symmetric polarization-insensitive directional couplers fabricated
  by femtosecond laser writing},\ }\href@noop {} {\bibfield  {journal}
  {\bibinfo  {journal} {Optics Express}\ }\textbf {\bibinfo {volume} {26}},\
  \bibinfo {pages} {15101} (\bibinfo {year} {2018})}\BibitemShut {NoStop}%
\bibitem [{\citenamefont {Vest}\ \emph {et~al.}(2014)\citenamefont {Vest},
  \citenamefont {Rau}, \citenamefont {Fuchs}, \citenamefont {Corrielli},
  \citenamefont {Weier}, \citenamefont {Nauerth}, \citenamefont {Crespi},
  \citenamefont {Osellame},\ and\ \citenamefont {Weinfurter}}]{vest2014design}%
  \BibitemOpen
  \bibfield  {author} {\bibinfo {author} {\bibfnamefont {G.}~\bibnamefont
  {Vest}}, \bibinfo {author} {\bibfnamefont {M.}~\bibnamefont {Rau}}, \bibinfo
  {author} {\bibfnamefont {L.}~\bibnamefont {Fuchs}}, \bibinfo {author}
  {\bibfnamefont {G.}~\bibnamefont {Corrielli}}, \bibinfo {author}
  {\bibfnamefont {H.}~\bibnamefont {Weier}}, \bibinfo {author} {\bibfnamefont
  {S.}~\bibnamefont {Nauerth}}, \bibinfo {author} {\bibfnamefont
  {A.}~\bibnamefont {Crespi}}, \bibinfo {author} {\bibfnamefont
  {R.}~\bibnamefont {Osellame}},\ and\ \bibinfo {author} {\bibfnamefont
  {H.}~\bibnamefont {Weinfurter}},\ }\bibfield  {title} {\bibinfo {title}
  {Design and evaluation of a handheld quantum key distribution sender
  module},\ }\href@noop {} {\bibfield  {journal} {\bibinfo  {journal} {IEEE
  journal of selected topics in quantum electronics}\ }\textbf {\bibinfo
  {volume} {21}},\ \bibinfo {pages} {131} (\bibinfo {year} {2014})}\BibitemShut
  {NoStop}%
\bibitem [{\citenamefont {Vogl}\ \emph {et~al.}(2017)\citenamefont {Vogl},
  \citenamefont {Lu},\ and\ \citenamefont {Lam}}]{0022-3727-50-29-295101}%
  \BibitemOpen
  \bibfield  {author} {\bibinfo {author} {\bibfnamefont {T.}~\bibnamefont
  {Vogl}}, \bibinfo {author} {\bibfnamefont {Y.}~\bibnamefont {Lu}},\ and\
  \bibinfo {author} {\bibfnamefont {P.~K.}\ \bibnamefont {Lam}},\ }\bibfield
  {title} {\bibinfo {title} {Room temperature single photon source using
  fiber-integrated hexagonal boron nitride},\ }\href
  {http://stacks.iop.org/0022-3727/50/i=29/a=295101} {\bibfield  {journal}
  {\bibinfo  {journal} {J. Phys. D: Appl. Phys.}\ }\textbf {\bibinfo {volume}
  {50}},\ \bibinfo {pages} {295101} (\bibinfo {year} {2017})}\BibitemShut
  {NoStop}%
\bibitem [{\citenamefont {NASA}(2013)}]{GSFC-STD-7000}%
  \BibitemOpen
  \bibfield  {author} {\bibinfo {author} {\bibnamefont {NASA}},\ }\href@noop {}
  {\bibinfo {title} {{General Environmental Verification Standard
  GSFC-STD-7000}}} (\bibinfo {year} {2013}),\ \bibinfo {note}
  {\href{https://standards.nasa.gov/standard/gsfc/gsfc-std-7000}{https://standards.nasa.gov/standard/gsfc/gsfc-std-7000}}\BibitemShut
  {NoStop}%
\bibitem [{\citenamefont {Tan}\ \emph {et~al.}(2013)\citenamefont {Tan},
  \citenamefont {Chandrasekara}, \citenamefont {Cheng},\ and\ \citenamefont
  {Ling}}]{Tan:13}%
  \BibitemOpen
  \bibfield  {author} {\bibinfo {author} {\bibfnamefont {Y.~C.}\ \bibnamefont
  {Tan}}, \bibinfo {author} {\bibfnamefont {R.}~\bibnamefont {Chandrasekara}},
  \bibinfo {author} {\bibfnamefont {C.}~\bibnamefont {Cheng}},\ and\ \bibinfo
  {author} {\bibfnamefont {A.}~\bibnamefont {Ling}},\ }\bibfield  {title}
  {\bibinfo {title} {Silicon avalanche photodiode operation and lifetime
  analysis for small satellites},\ }\href
  {https://doi.org/10.1364/OE.21.016946} {\bibfield  {journal} {\bibinfo
  {journal} {Opt. Express}\ }\textbf {\bibinfo {volume} {21}},\ \bibinfo
  {pages} {16946} (\bibinfo {year} {2013})}\BibitemShut {NoStop}%
\bibitem [{\citenamefont {Vogl}\ \emph
  {et~al.}(2019{\natexlab{a}})\citenamefont {Vogl}, \citenamefont {Sripathy},
  \citenamefont {Sharma}, \citenamefont {Reddy}, \citenamefont {Sullivan},
  \citenamefont {Machacek}, \citenamefont {Zhang}, \citenamefont {Karouta},
  \citenamefont {Buchler}, \citenamefont {Doherty}, \citenamefont {Lu},\ and\
  \citenamefont {Lam}}]{10.1038/s41467-019-09219-5}%
  \BibitemOpen
  \bibfield  {author} {\bibinfo {author} {\bibfnamefont {T.}~\bibnamefont
  {Vogl}}, \bibinfo {author} {\bibfnamefont {K.}~\bibnamefont {Sripathy}},
  \bibinfo {author} {\bibfnamefont {A.}~\bibnamefont {Sharma}}, \bibinfo
  {author} {\bibfnamefont {P.}~\bibnamefont {Reddy}}, \bibinfo {author}
  {\bibfnamefont {J.}~\bibnamefont {Sullivan}}, \bibinfo {author}
  {\bibfnamefont {J.~R.}\ \bibnamefont {Machacek}}, \bibinfo {author}
  {\bibfnamefont {L.}~\bibnamefont {Zhang}}, \bibinfo {author} {\bibfnamefont
  {F.}~\bibnamefont {Karouta}}, \bibinfo {author} {\bibfnamefont {B.~C.}\
  \bibnamefont {Buchler}}, \bibinfo {author} {\bibfnamefont {M.~W.}\
  \bibnamefont {Doherty}}, \bibinfo {author} {\bibfnamefont {Y.}~\bibnamefont
  {Lu}},\ and\ \bibinfo {author} {\bibfnamefont {P.~K.}\ \bibnamefont {Lam}},\
  }\bibfield  {title} {\bibinfo {title} {Radiation tolerance of two-dimensional
  material-based devices for space applications},\ }\href
  {https://doi.org/10.1038/s41467-019-09219-5} {\bibfield  {journal} {\bibinfo
  {journal} {Nat. Commun.}\ }\textbf {\bibinfo {volume} {10}},\ \bibinfo
  {pages} {1202} (\bibinfo {year} {2019}{\natexlab{a}})}\BibitemShut {NoStop}%
\bibitem [{\citenamefont {Vogl}\ \emph
  {et~al.}(2019{\natexlab{b}})\citenamefont {Vogl}, \citenamefont {Lecamwasam},
  \citenamefont {Buchler}, \citenamefont {Lu},\ and\ \citenamefont
  {Lam}}]{doi:10.1021/acsphotonics.9b00314}%
  \BibitemOpen
  \bibfield  {author} {\bibinfo {author} {\bibfnamefont {T.}~\bibnamefont
  {Vogl}}, \bibinfo {author} {\bibfnamefont {R.}~\bibnamefont {Lecamwasam}},
  \bibinfo {author} {\bibfnamefont {B.~C.}\ \bibnamefont {Buchler}}, \bibinfo
  {author} {\bibfnamefont {Y.}~\bibnamefont {Lu}},\ and\ \bibinfo {author}
  {\bibfnamefont {P.~K.}\ \bibnamefont {Lam}},\ }\bibfield  {title} {\bibinfo
  {title} {{Compact Cavity-Enhanced Single-Photon Generation with Hexagonal
  Boron Nitride}},\ }\href {https://doi.org/10.1021/acsphotonics.9b00314}
  {\bibfield  {journal} {\bibinfo  {journal} {ACS Photonics}\ }\textbf
  {\bibinfo {volume} {6}},\ \bibinfo {pages} {1955} (\bibinfo {year}
  {2019}{\natexlab{b}})}\BibitemShut {NoStop}%
\bibitem [{\citenamefont {{European Space Agency}}()}]{spenvis}%
  \BibitemOpen
  \bibfield  {author} {\bibinfo {author} {\bibnamefont {{European Space
  Agency}}},\ }\href@noop {} {\bibinfo {title} {{The Space Environment
  Information System}}},\ \bibinfo {note}
  {\href{http://www.spenvis.oma.be/}{http://www.spenvis.oma.be/} accessed
  22/02/2020}\BibitemShut {NoStop}%
\bibitem [{\citenamefont {Arriola}\ \emph {et~al.}(2013)\citenamefont
  {Arriola}, \citenamefont {Gross}, \citenamefont {Jovanovic}, \citenamefont
  {Charles}, \citenamefont {Tuthill}, \citenamefont {Olaizola}, \citenamefont
  {Fuerbach},\ and\ \citenamefont {Withford}}]{arriola2013}%
  \BibitemOpen
  \bibfield  {author} {\bibinfo {author} {\bibfnamefont {A.}~\bibnamefont
  {Arriola}}, \bibinfo {author} {\bibfnamefont {S.}~\bibnamefont {Gross}},
  \bibinfo {author} {\bibfnamefont {N.}~\bibnamefont {Jovanovic}}, \bibinfo
  {author} {\bibfnamefont {N.}~\bibnamefont {Charles}}, \bibinfo {author}
  {\bibfnamefont {P.~G.}\ \bibnamefont {Tuthill}}, \bibinfo {author}
  {\bibfnamefont {S.~M.}\ \bibnamefont {Olaizola}}, \bibinfo {author}
  {\bibfnamefont {A.}~\bibnamefont {Fuerbach}},\ and\ \bibinfo {author}
  {\bibfnamefont {M.~J.}\ \bibnamefont {Withford}},\ }\bibfield  {title}
  {\bibinfo {title} {Low bend loss waveguides enable compact, efficient 3d
  photonic chips},\ }\href@noop {} {\bibfield  {journal} {\bibinfo  {journal}
  {Optics Express}\ }\textbf {\bibinfo {volume} {21}},\ \bibinfo {pages} {2978}
  (\bibinfo {year} {2013})}\BibitemShut {NoStop}%
\bibitem [{\citenamefont {Ziegler}\ \emph {et~al.}(2010)\citenamefont
  {Ziegler}, \citenamefont {Ziegler},\ and\ \citenamefont
  {Biersack}}]{10.1016/j.nimb.2010.02.091}%
  \BibitemOpen
  \bibfield  {author} {\bibinfo {author} {\bibfnamefont {J.~F.}\ \bibnamefont
  {Ziegler}}, \bibinfo {author} {\bibfnamefont {M.}~\bibnamefont {Ziegler}},\
  and\ \bibinfo {author} {\bibfnamefont {J.}~\bibnamefont {Biersack}},\
  }\bibfield  {title} {\bibinfo {title} {{SRIM - The stopping and range of ions
  in matter}},\ }\href {https://doi.org/10.1016/j.nimb.2010.02.091} {\bibfield
  {journal} {\bibinfo  {journal} {Nucl. Instr. Meth. Phys. Res. B}\ }\textbf
  {\bibinfo {volume} {268}},\ \bibinfo {pages} {1818 } (\bibinfo {year}
  {2010})}\BibitemShut {NoStop}%
\bibitem [{\citenamefont {Chu}\ \emph {et~al.}()\citenamefont {Chu},
  \citenamefont {Ekstr\"om},\ and\ \citenamefont {Firestone}}]{toi}%
  \BibitemOpen
  \bibfield  {author} {\bibinfo {author} {\bibfnamefont {S.~Y.~F.}\
  \bibnamefont {Chu}}, \bibinfo {author} {\bibfnamefont {L.~P.}\ \bibnamefont
  {Ekstr\"om}},\ and\ \bibinfo {author} {\bibfnamefont {R.~B.}\ \bibnamefont
  {Firestone}},\ }\href@noop {} {\bibinfo {title} {{WWW Table of Radioactive
  Isotopes}}},\ \bibinfo {note}
  {\href{http://nucleardata.nuclear.lu.se/nucleardata/toi/}{http://nucleardata.nuclear.lu.se/nucleardata/toi/}
  database version 28/02/1999}\BibitemShut {NoStop}%
\bibitem [{\citenamefont {Osellame}\ \emph {et~al.}(2004)\citenamefont
  {Osellame}, \citenamefont {Chiodo}, \citenamefont {Della~Valle},
  \citenamefont {Taccheo}, \citenamefont {Ramponi}, \citenamefont {Cerullo},
  \citenamefont {Killi}, \citenamefont {Morgner}, \citenamefont {Lederer},\
  and\ \citenamefont {Kopf}}]{osellame2004optical}%
  \BibitemOpen
  \bibfield  {author} {\bibinfo {author} {\bibfnamefont {R.}~\bibnamefont
  {Osellame}}, \bibinfo {author} {\bibfnamefont {N.}~\bibnamefont {Chiodo}},
  \bibinfo {author} {\bibfnamefont {G.}~\bibnamefont {Della~Valle}}, \bibinfo
  {author} {\bibfnamefont {S.}~\bibnamefont {Taccheo}}, \bibinfo {author}
  {\bibfnamefont {R.}~\bibnamefont {Ramponi}}, \bibinfo {author} {\bibfnamefont
  {G.}~\bibnamefont {Cerullo}}, \bibinfo {author} {\bibfnamefont
  {A.}~\bibnamefont {Killi}}, \bibinfo {author} {\bibfnamefont
  {U.}~\bibnamefont {Morgner}}, \bibinfo {author} {\bibfnamefont
  {M.}~\bibnamefont {Lederer}},\ and\ \bibinfo {author} {\bibfnamefont
  {D.}~\bibnamefont {Kopf}},\ }\bibfield  {title} {\bibinfo {title} {Optical
  waveguide writing with a diode-pumped femtosecond oscillator},\ }\href@noop
  {} {\bibfield  {journal} {\bibinfo  {journal} {Optics letters}\ }\textbf
  {\bibinfo {volume} {29}},\ \bibinfo {pages} {1900} (\bibinfo {year}
  {2004})}\BibitemShut {NoStop}%
\bibitem [{\citenamefont {Kataoka}\ \emph {et~al.}(2010)\citenamefont
  {Kataoka}, \citenamefont {Toizumi}, \citenamefont {Nakamori}, \citenamefont
  {Yatsu}, \citenamefont {Tsubuku}, \citenamefont {Kuramoto}, \citenamefont
  {Enomoto}, \citenamefont {Usui}, \citenamefont {Kawai}, \citenamefont
  {Ashida}, \citenamefont {Omagari}, \citenamefont {Fujihashi}, \citenamefont
  {Inagawa}, \citenamefont {Miura}, \citenamefont {Konda}, \citenamefont
  {Miyashita}, \citenamefont {Matsunaga}, \citenamefont {Ishikawa},
  \citenamefont {Matsunaga},\ and\ \citenamefont
  {Kawabata}}]{doi:10.1029/2009JA014699}%
  \BibitemOpen
  \bibfield  {author} {\bibinfo {author} {\bibfnamefont {J.}~\bibnamefont
  {Kataoka}}, \bibinfo {author} {\bibfnamefont {T.}~\bibnamefont {Toizumi}},
  \bibinfo {author} {\bibfnamefont {T.}~\bibnamefont {Nakamori}}, \bibinfo
  {author} {\bibfnamefont {Y.}~\bibnamefont {Yatsu}}, \bibinfo {author}
  {\bibfnamefont {Y.}~\bibnamefont {Tsubuku}}, \bibinfo {author} {\bibfnamefont
  {Y.}~\bibnamefont {Kuramoto}}, \bibinfo {author} {\bibfnamefont
  {T.}~\bibnamefont {Enomoto}}, \bibinfo {author} {\bibfnamefont
  {R.}~\bibnamefont {Usui}}, \bibinfo {author} {\bibfnamefont {N.}~\bibnamefont
  {Kawai}}, \bibinfo {author} {\bibfnamefont {H.}~\bibnamefont {Ashida}},
  \bibinfo {author} {\bibfnamefont {K.}~\bibnamefont {Omagari}}, \bibinfo
  {author} {\bibfnamefont {K.}~\bibnamefont {Fujihashi}}, \bibinfo {author}
  {\bibfnamefont {S.}~\bibnamefont {Inagawa}}, \bibinfo {author} {\bibfnamefont
  {Y.}~\bibnamefont {Miura}}, \bibinfo {author} {\bibfnamefont
  {Y.}~\bibnamefont {Konda}}, \bibinfo {author} {\bibfnamefont
  {N.}~\bibnamefont {Miyashita}}, \bibinfo {author} {\bibfnamefont
  {S.}~\bibnamefont {Matsunaga}}, \bibinfo {author} {\bibfnamefont
  {Y.}~\bibnamefont {Ishikawa}}, \bibinfo {author} {\bibfnamefont
  {Y.}~\bibnamefont {Matsunaga}},\ and\ \bibinfo {author} {\bibfnamefont
  {N.}~\bibnamefont {Kawabata}},\ }\bibfield  {title} {\bibinfo {title}
  {{In-orbit performance of avalanche photodiode as radiation detector on board
  the picosatellite Cute-1.7+APD II}},\ }\href
  {https://doi.org/10.1029/2009JA014699} {\bibfield  {journal} {\bibinfo
  {journal} {Journal of Geophysical Research: Space Physics}\ }\textbf
  {\bibinfo {volume} {115}} (\bibinfo {year} {2010})}\BibitemShut {NoStop}%
\bibitem [{\citenamefont {Labadie}\ \emph {et~al.}(2016)\citenamefont
  {Labadie}, \citenamefont {Berger}, \citenamefont {Cvetojevic}, \citenamefont
  {Haynes}, \citenamefont {Harris}, \citenamefont {Jovanovic}, \citenamefont
  {Lacour}, \citenamefont {Martin}, \citenamefont {Minardi}, \citenamefont
  {Perrin} \emph {et~al.}}]{labadie2016astronomical}%
  \BibitemOpen
  \bibfield  {author} {\bibinfo {author} {\bibfnamefont {L.}~\bibnamefont
  {Labadie}}, \bibinfo {author} {\bibfnamefont {J.-P.}\ \bibnamefont {Berger}},
  \bibinfo {author} {\bibfnamefont {N.}~\bibnamefont {Cvetojevic}}, \bibinfo
  {author} {\bibfnamefont {R.}~\bibnamefont {Haynes}}, \bibinfo {author}
  {\bibfnamefont {R.}~\bibnamefont {Harris}}, \bibinfo {author} {\bibfnamefont
  {N.}~\bibnamefont {Jovanovic}}, \bibinfo {author} {\bibfnamefont
  {S.}~\bibnamefont {Lacour}}, \bibinfo {author} {\bibfnamefont
  {G.}~\bibnamefont {Martin}}, \bibinfo {author} {\bibfnamefont
  {S.}~\bibnamefont {Minardi}}, \bibinfo {author} {\bibfnamefont
  {G.}~\bibnamefont {Perrin}}, \emph {et~al.},\ }\bibfield  {title} {\bibinfo
  {title} {Astronomical photonics in the context of infrared interferometry and
  high-resolution spectroscopy},\ }in\ \href@noop {} {\emph {\bibinfo
  {booktitle} {Optical and Infrared Interferometry and Imaging V}}},\ Vol.\
  \bibinfo {volume} {9907}\ (\bibinfo {organization} {International Society for
  Optics and Photonics},\ \bibinfo {year} {2016})\ p.\ \bibinfo {pages}
  {990718}\BibitemShut {NoStop}%
\bibitem [{\citenamefont {Diener}\ \emph {et~al.}(2017)\citenamefont {Diener},
  \citenamefont {Tepper}, \citenamefont {Labadie}, \citenamefont {Pertsch},
  \citenamefont {Nolte},\ and\ \citenamefont {Minardi}}]{diener2017towards}%
  \BibitemOpen
  \bibfield  {author} {\bibinfo {author} {\bibfnamefont {R.}~\bibnamefont
  {Diener}}, \bibinfo {author} {\bibfnamefont {J.}~\bibnamefont {Tepper}},
  \bibinfo {author} {\bibfnamefont {L.}~\bibnamefont {Labadie}}, \bibinfo
  {author} {\bibfnamefont {T.}~\bibnamefont {Pertsch}}, \bibinfo {author}
  {\bibfnamefont {S.}~\bibnamefont {Nolte}},\ and\ \bibinfo {author}
  {\bibfnamefont {S.}~\bibnamefont {Minardi}},\ }\bibfield  {title} {\bibinfo
  {title} {{Towards 3D-photonic, multi-telescope beam combiners for
  mid-infrared astrointerferometry}},\ }\href@noop {} {\bibfield  {journal}
  {\bibinfo  {journal} {Optics Express}\ }\textbf {\bibinfo {volume} {25}},\
  \bibinfo {pages} {19262} (\bibinfo {year} {2017})}\BibitemShut {NoStop}%
\bibitem [{\citenamefont {Norris}\ \emph {et~al.}(2020)\citenamefont {Norris},
  \citenamefont {Cvetojevic}, \citenamefont {Lagadec}, \citenamefont
  {Jovanovic}, \citenamefont {Gross}, \citenamefont {Arriola}, \citenamefont
  {Gretzinger}, \citenamefont {Martinod}, \citenamefont {Guyon}, \citenamefont
  {Lozi} \emph {et~al.}}]{norris2020first}%
  \BibitemOpen
  \bibfield  {author} {\bibinfo {author} {\bibfnamefont {B.~R.}\ \bibnamefont
  {Norris}}, \bibinfo {author} {\bibfnamefont {N.}~\bibnamefont {Cvetojevic}},
  \bibinfo {author} {\bibfnamefont {T.}~\bibnamefont {Lagadec}}, \bibinfo
  {author} {\bibfnamefont {N.}~\bibnamefont {Jovanovic}}, \bibinfo {author}
  {\bibfnamefont {S.}~\bibnamefont {Gross}}, \bibinfo {author} {\bibfnamefont
  {A.}~\bibnamefont {Arriola}}, \bibinfo {author} {\bibfnamefont
  {T.}~\bibnamefont {Gretzinger}}, \bibinfo {author} {\bibfnamefont {M.-A.}\
  \bibnamefont {Martinod}}, \bibinfo {author} {\bibfnamefont {O.}~\bibnamefont
  {Guyon}}, \bibinfo {author} {\bibfnamefont {J.}~\bibnamefont {Lozi}}, \emph
  {et~al.},\ }\bibfield  {title} {\bibinfo {title} {First on-sky demonstration
  of an integrated-photonic nulling interferometer: the glint instrument},\
  }\href@noop {} {\bibfield  {journal} {\bibinfo  {journal} {Monthly Notices of
  the Royal Astronomical Society}\ }\textbf {\bibinfo {volume} {491}},\
  \bibinfo {pages} {4180} (\bibinfo {year} {2020})}\BibitemShut {NoStop}%
\end{thebibliography}
\providecommand{\noopsort}[1]{}\providecommand{\singleletter}[1]{#1}%

\clearpage
\onecolumngrid
\renewcommand\thesection{S\arabic{section}}
\setcounter{section}{0}
\renewcommand\thetable{S\arabic{table}}
\setcounter{table}{0}
\renewcommand\thefigure{S\arabic{figure}}
\setcounter{figure}{0}
\pagenumbering{arabic}
\renewcommand*{\thepage}{S\arabic{page}}
\section*{Supplementary Materials}
\section{Ion-target interaction}
\begin{figure*}[hbt]
\centerline{\includegraphics[width=\linewidth]{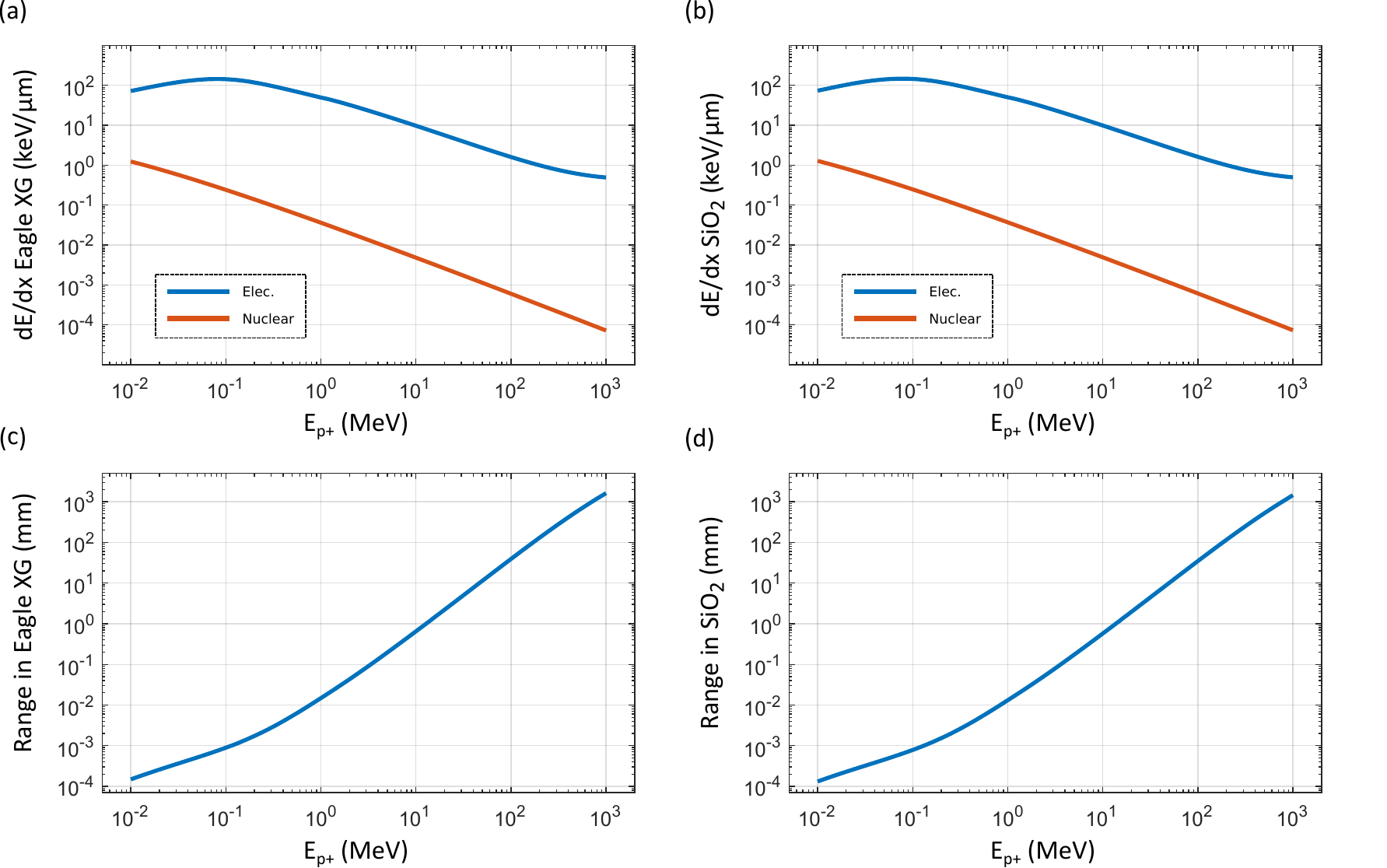}}
\caption{Stopping power and projected range. (a, b) Nuclear and electronic stopping power of protons in Eagle XG and \ce{SiO2}, respectively. (c, d) Projected range of protons in Eagle XG and \ce{SiO2}, respectively. Note that $\gamma$-rays are not attenuated in the glass, at least not significantly within the thickness of the glass chips.} 
\label{figSI1}
\end{figure*}

\begin{figure}[hbt]
\centering
\includegraphics[width=0.6\linewidth]{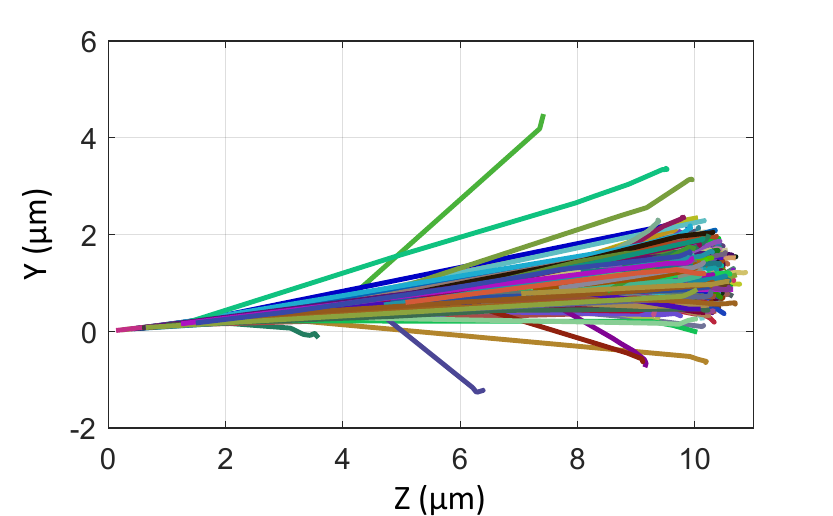}
\caption{Monte Carlo simulations of 500 proton trajectories in Eagle XG. The protons have an energy of 770 keV and the angle of incidence is 7$^\circ$ (see Main Text). The trajectories are terminated once each proton is stopped. At the proton energy of 770 keV, most of the protons are stopped at a depth of 10 $\mu$m, the depth at which some of our waveguides are buried.} 
\label{figSI2}
\end{figure}

To study the interaction of the protons with the glass, we used the SRIM (Stopping and Range of Ions in Matter) code \cite{10.1016/j.nimb.2010.02.091}. While the exact composition of Eagle XG (EXG) remains a trade secret, a typical composition is 55.0\,\% \ce{SiO2}, 7.0\,\% \ce{B2O3}, 10.4\,\% \ce{Al2O3}, 21.0\,\% \ce{CaO}, 1.0\,\% \ce{SrO}, and 5.6\,\% \ce{MgO}. The density is $\rho=2.38$ g/cm$^3$. Using SRIM we calculate the electronic and nuclear stopping power of protons in EXG and compare the results to 100\,\% \ce{SiO2} with a density of $\rho=2.65$ g/cm$^3$ as a reference (see figure \ref{figSI1}(a) for EXG and (b) for \ce{SiO2}). The stopping power is the proton energy loss as they travel through the material, which allows us to calculate the projected proton range into the materials (see figure \ref{figSI1}(c) for EXG and (d) for \ce{SiO2}). Protons with an energy of 770 keV have a projected range of 10 $\mu$m, while 3 MeV protons can penetrate much deeper. As expected, there are only little differences between EXG and pure \ce{SiO2}. The stopping power decreasing with increasing energy means that higher-energy protons will simply pass through the waveguide, however, they will ionize and sputter atoms on their way. By implication this also means that lower-energy protons can deposit significantly more energy into the material. We also show a Monte Carlo simulation of 500 proton trajectories fired into the glass. The proton energy is 770 keV and the trajectories terminate when the ion is stopped (see figure \ref{figSI2}).

\section{Microscope inspection of the optical waveguides after radiation exposure}
\begin{figure*}[hbt]
\centerline{\includegraphics[width=\linewidth]{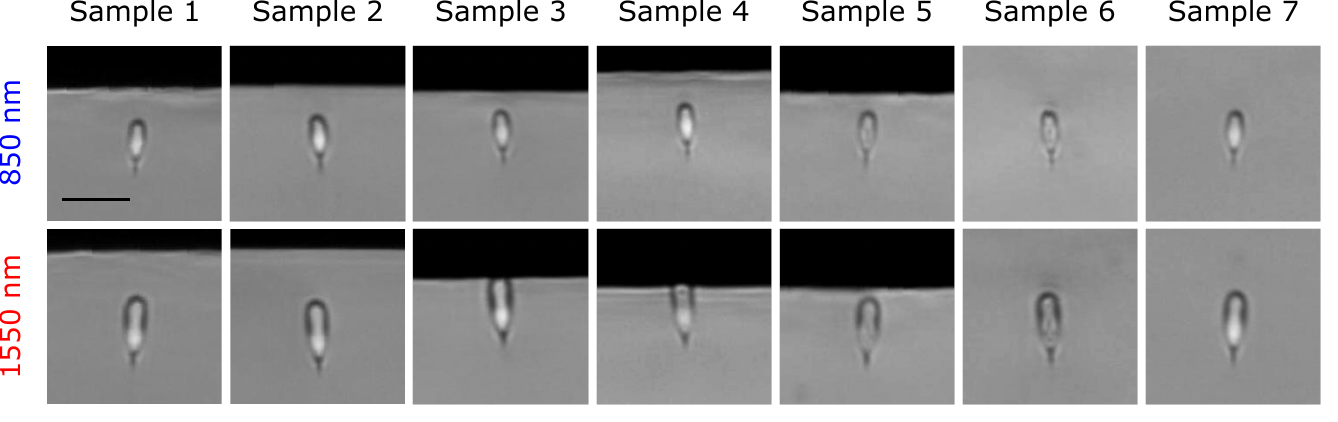}}
\caption{Microscope images of the waveguide cross sections measured after their irradiation. The scale bar is 10 $\mu$m.} 
\label{figSI3}
\end{figure*}

We report in figure \ref{figSI3} the results of the microscope inspection of the samples performed after the irradiation to $\gamma$-rays and protons. In particular, we show the details of the cross section of one waveguide working at 850 nm and one at 1550 nm for every sample. For the radiation exposure parameters of each sample we refer to the main text. From this inspection it is possible to appreciate that sample 5 and sample 6 clearly show a cross section degradation, not visible for all other samples. This observation is consistent with the increase of waveguide propagation losses measured on these samples.

\section{Bar transmission measurement: complete data set}
\begin{figure*}[hbt]
\centerline{\includegraphics[width=\linewidth]{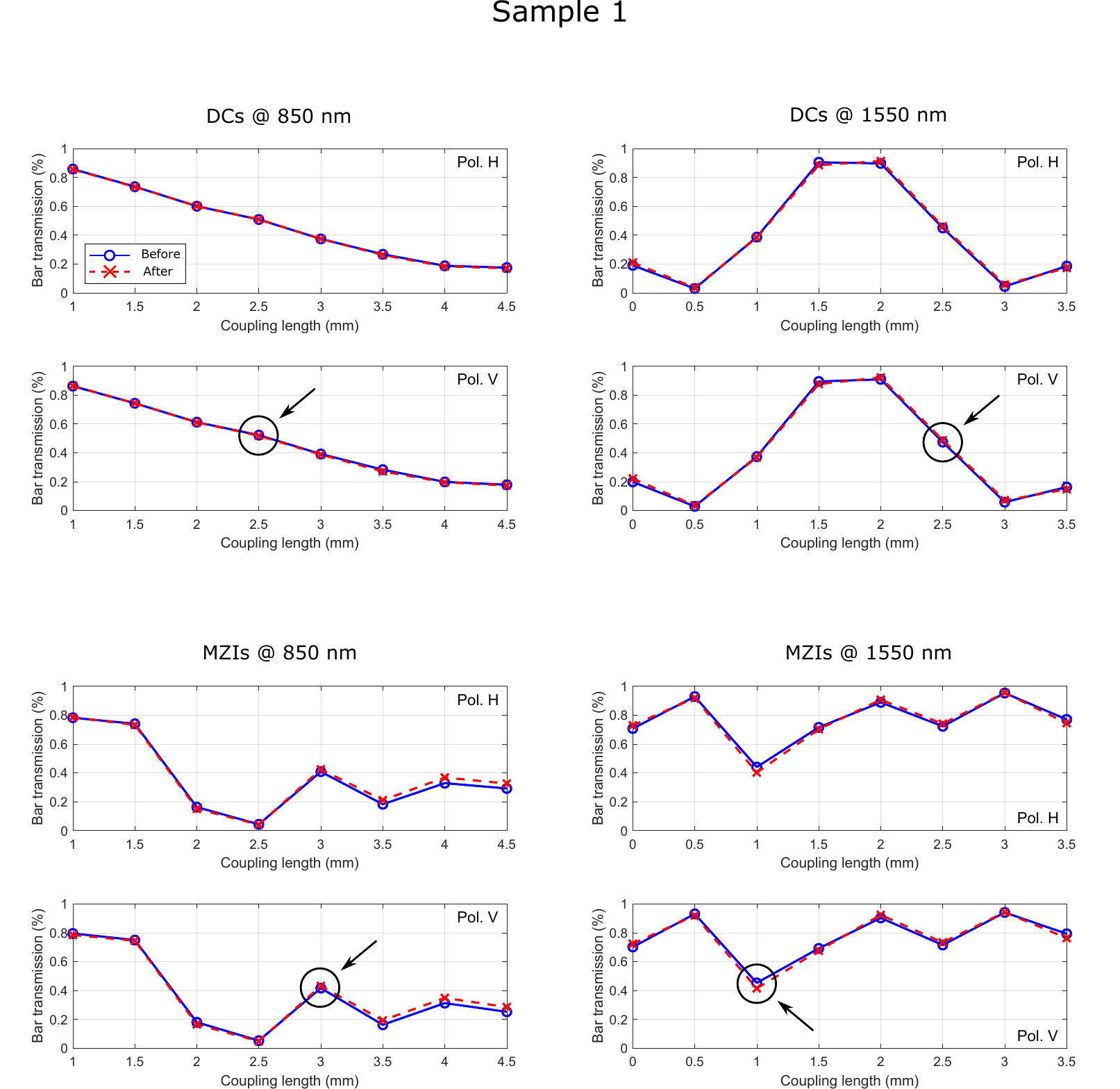}}
\caption{Complete data set of the measurements performed on the DCs and the MZIs of sample 1. The arrows indicate the devices highlighted in the main text.} 
\label{SI4s1}
\end{figure*}

\begin{figure*}[hbt]
\centerline{\includegraphics[width=\linewidth]{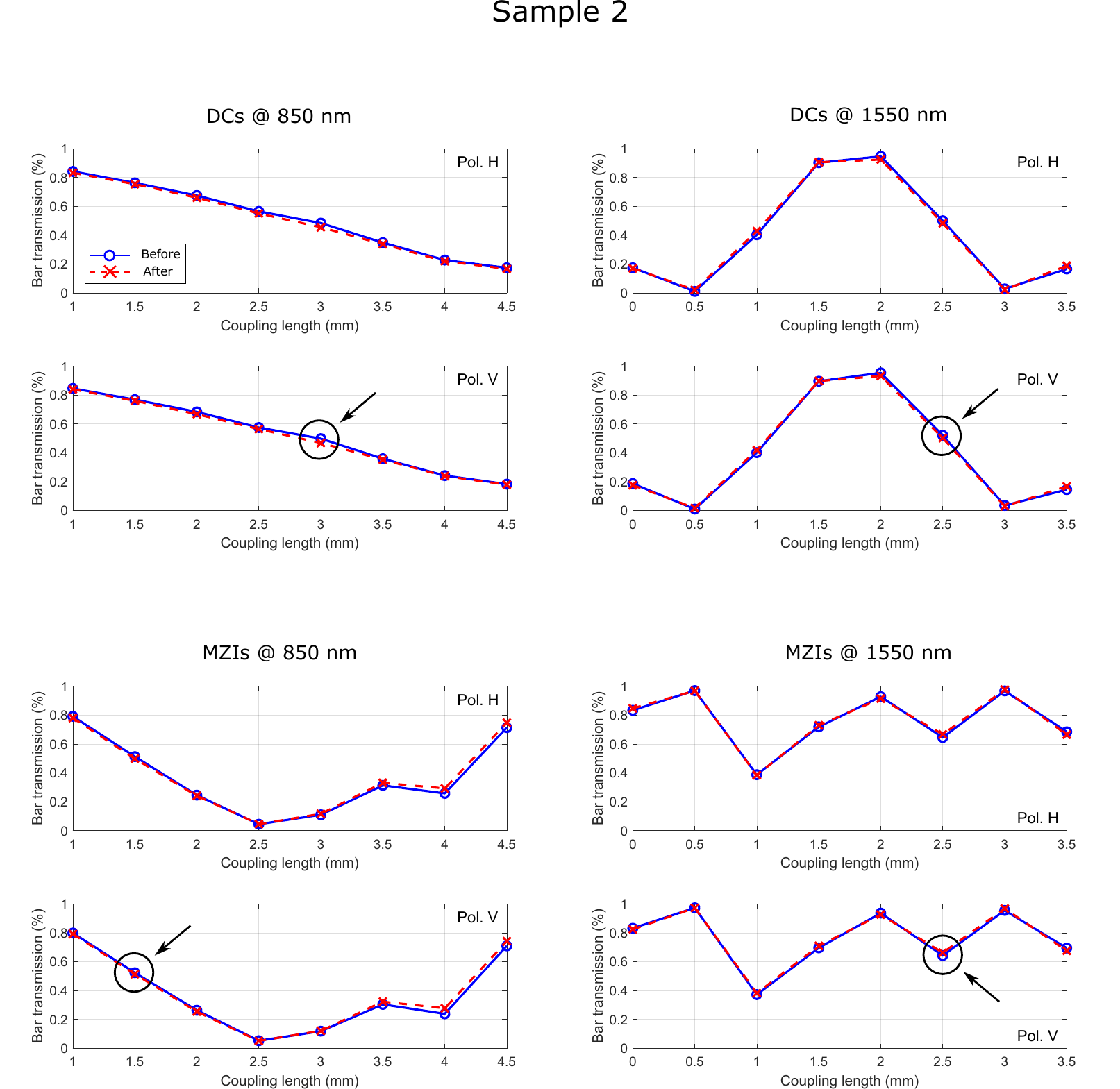}}
\caption{Complete data set of the measurements performed on the DCs and the MZIs of sample 2. The arrows indicate the devices highlighted in the main text.} 
\label{SI4s2}
\end{figure*}

\begin{figure*}[hbt]
\centerline{\includegraphics[width=\linewidth]{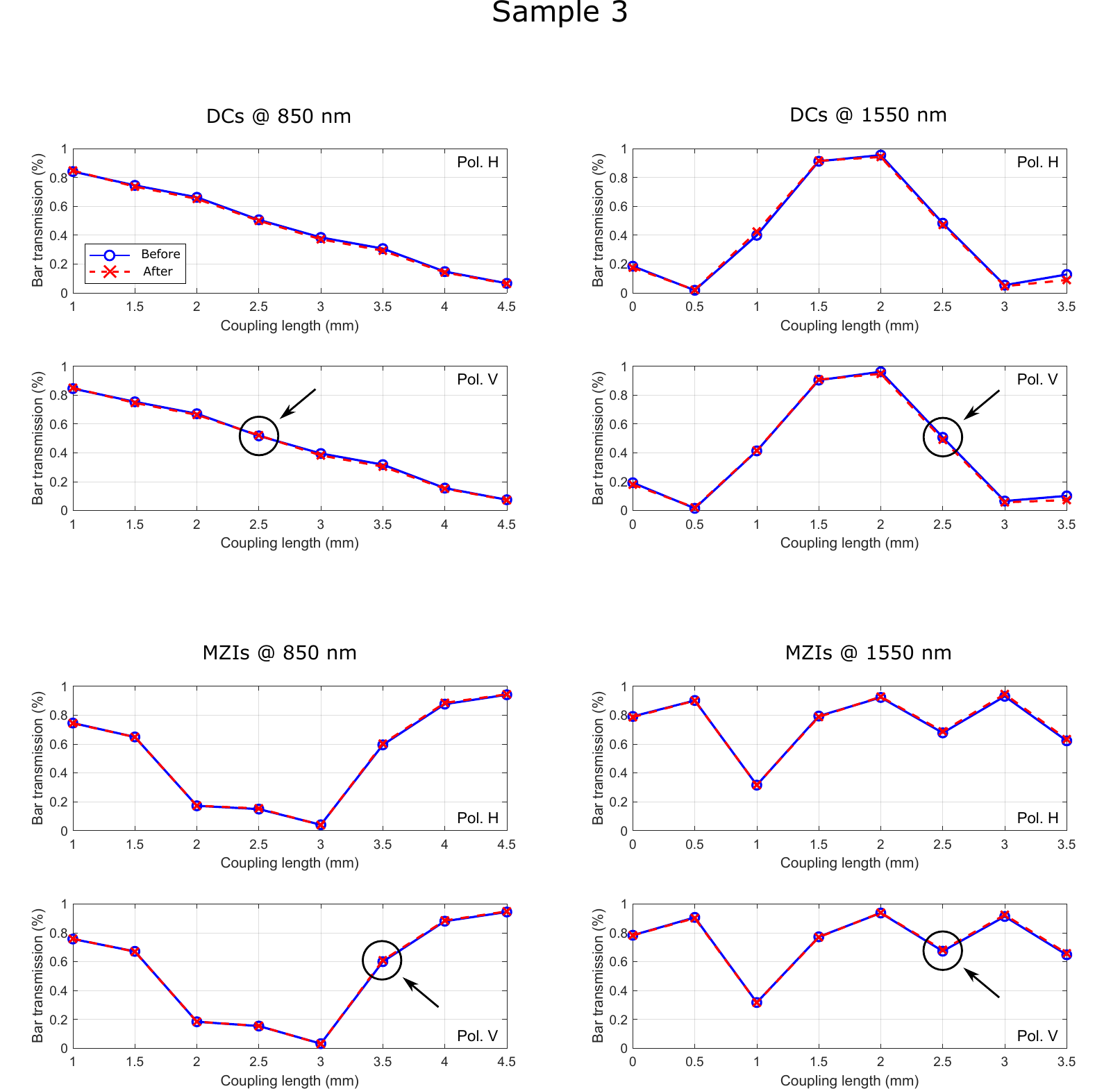}}
\caption{Complete data set of the measurements performed on the DCs and the MZIs of sample 3. The arrows indicate the devices highlighted in the main text.} 
\label{SI4s3}
\end{figure*}

\begin{figure*}[hbt]
\centerline{\includegraphics[width=\linewidth]{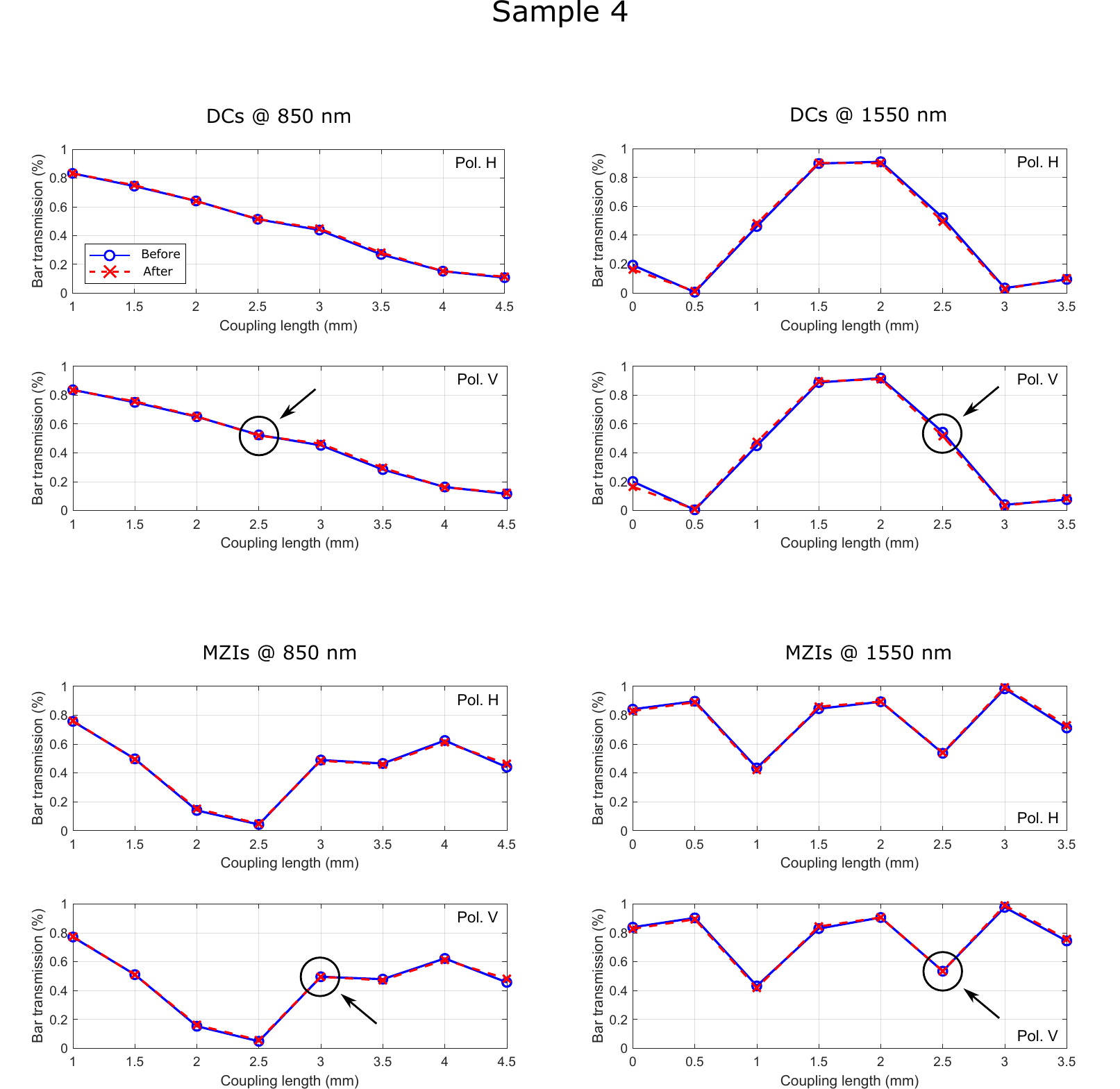}}
\caption{Complete data set of the measurements performed on the DCs and the MZIs of sample 4. The arrows indicate the devices highlighted in the main text.} 
\label{SI4s4}
\end{figure*}

\begin{figure*}[hbt]
\centerline{\includegraphics[width=\linewidth]{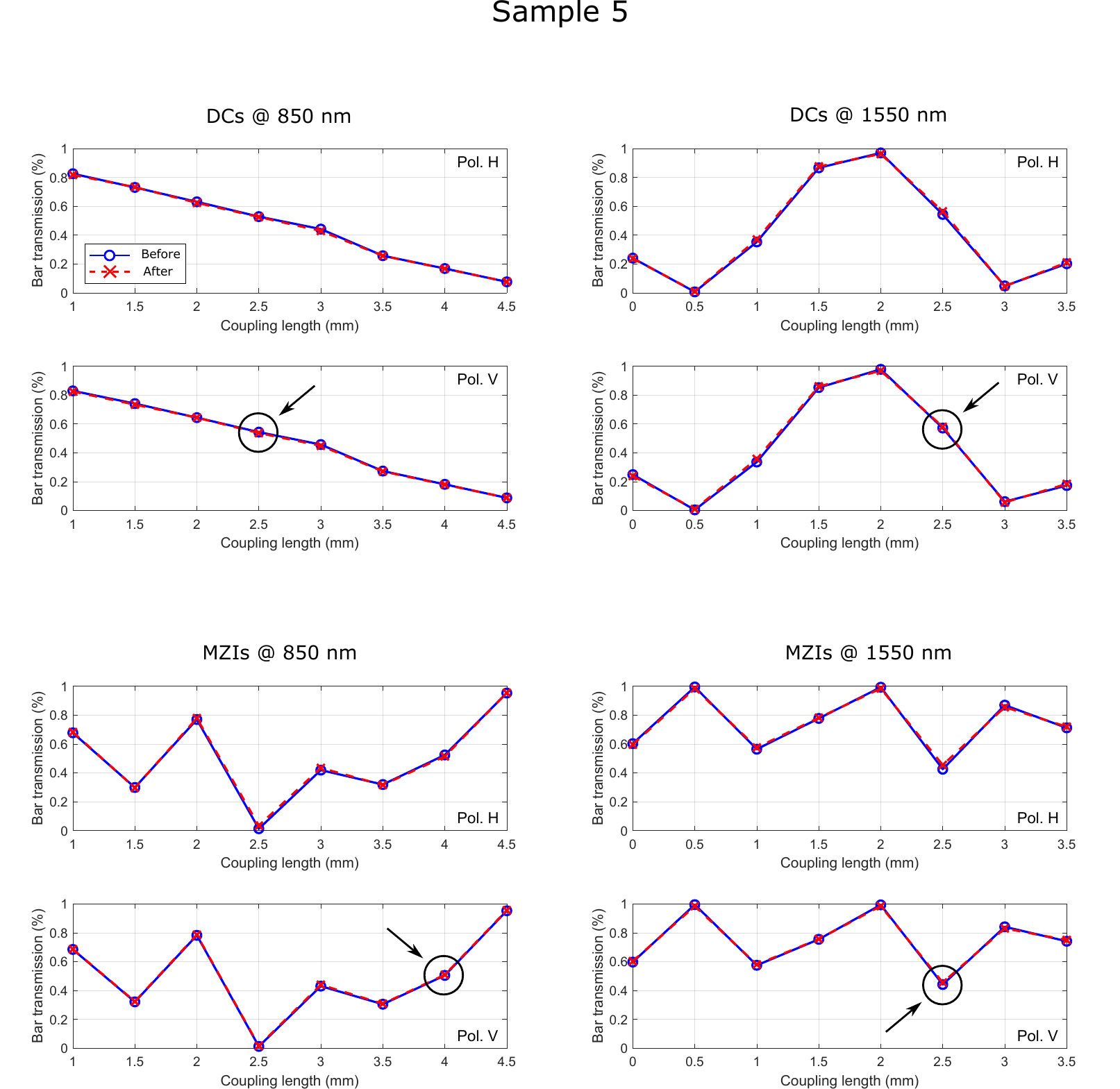}}
\caption{Complete data set of the measurements performed on the DCs and the MZIs of sample 5. The arrows indicate the devices highlighted in the main text.} 
\label{SI4s5}
\end{figure*}

\begin{figure*}[hbt]
\centerline{\includegraphics[width=\linewidth]{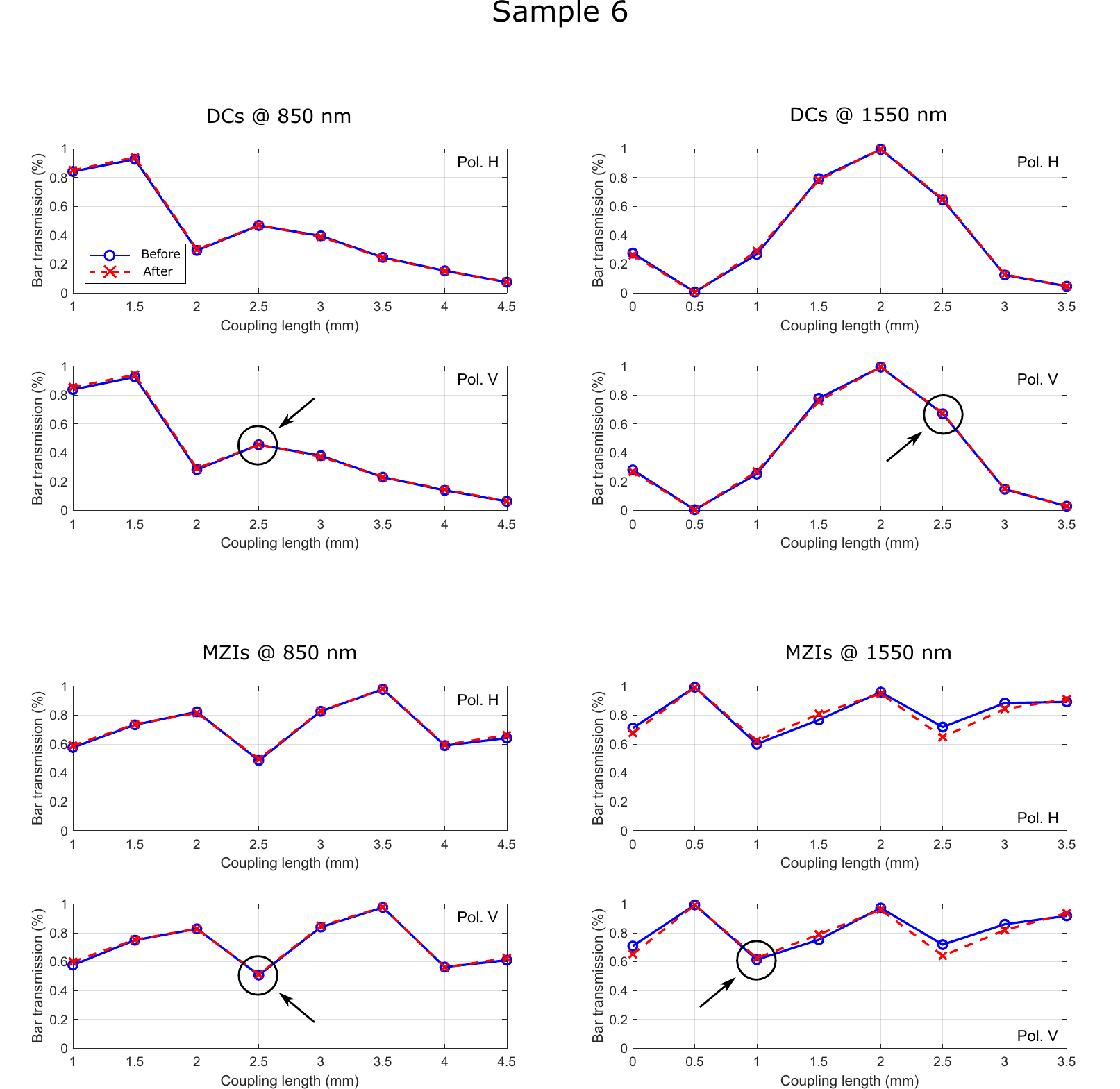}}
\caption{Complete data set of the measurements performed on the DCs and the MZIs of sample 6. The arrows indicate the devices highlighted in the main text.} 
\label{SI4s6}
\end{figure*}

\begin{figure*}[hbt]
\centerline{\includegraphics[width=\linewidth]{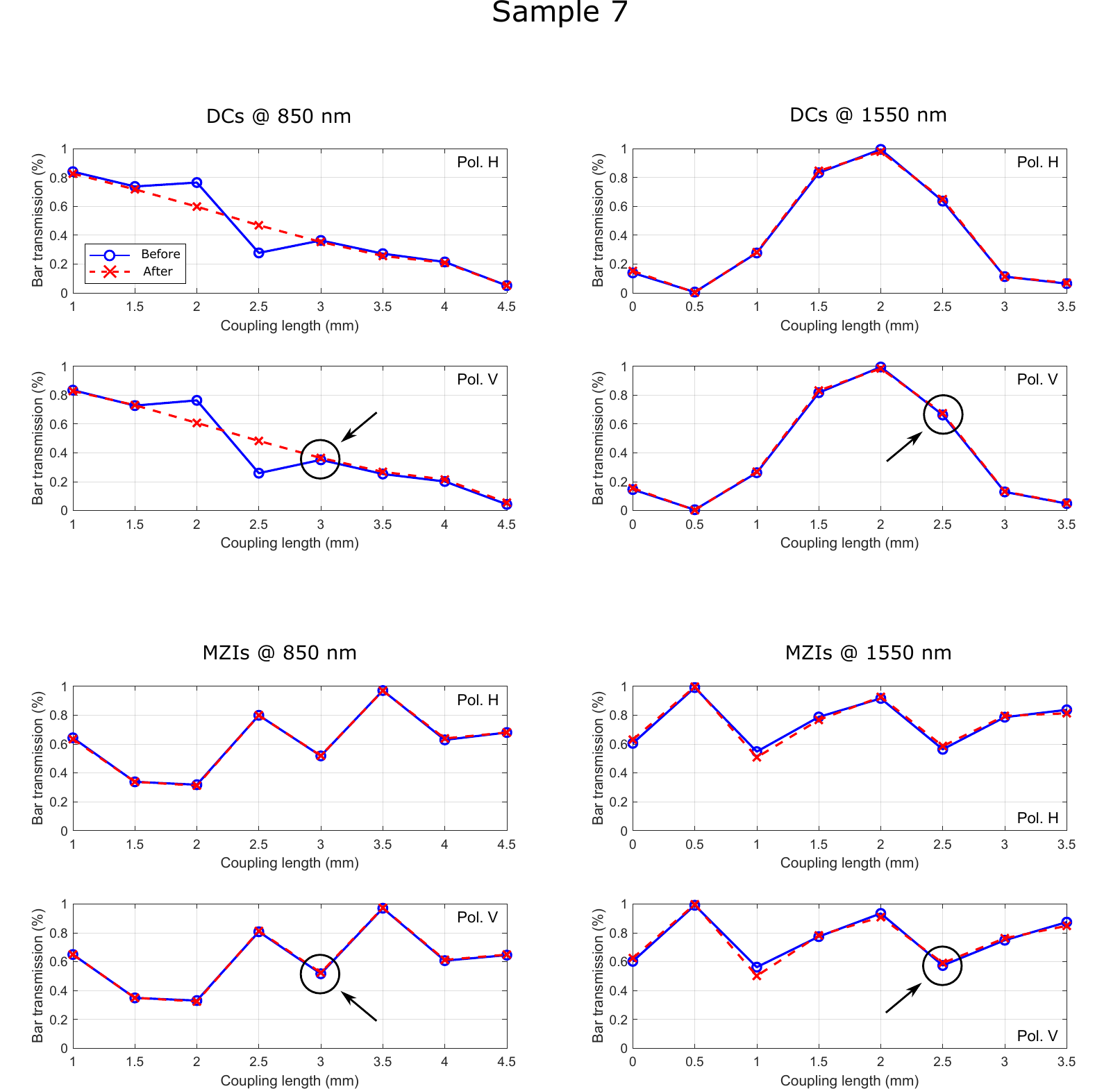}}
\caption{Complete data set of the measurements performed on the DCs and the MZIs of sample 7. The arrows indicate the devices highlighted in the main text.} 
\label{SI4s7}
\end{figure*}

In figures \ref{SI4s1}-\ref{SI4s7} we report the plots representing the complete data set of the splitting ratio measurements we performed on all Directional Couplers (DCs) and Mach-Zehnder Interferometers (MZIs). In these graphs, blue circles represent the data taken before the radiation exposure, while red crosses represent the data taken after it. The error bars are smaller than the data markers. It is worth highlighting that the values corresponding to the DCs of sample 7 (figure \ref{SI4s7}), operating at 850 nm and with interaction length of 2.5 mm and 3 mm present a strong variation before and after exposure. However, especially in light of the fact that we observed this effect only in sample 7, in just one of the two wavelength subset, and solely for two devices out of sixteen, we believe that it is not statistically relevant, and we attribute it to a mistake during the acquisition of the data before the irradiation.

\end{document}